\documentclass[5p,twocolumn]{elsart}

\usepackage{hyperref}
\usepackage[caption=false,font=footnotesize]{subfig}
\usepackage{microtype}
\usepackage{graphicx}
\usepackage{array}
\usepackage{multirow}
\usepackage{bold-extra}
\usepackage{balance}
\usepackage{booktabs}
\usepackage{color}
\usepackage{breakurl}

\usepackage{amsmath}

\usepackage{algorithm}
\usepackage{algpseudocode}

\usepackage{float}

\usepackage{soul}

\usepackage[final]{changes}

\begin{document}

\title{A Reinforcement Learning-based Link Quality Estimation Strategy for RPL and its Impact on Topology Management}

\author[addr1]{Emilio~Ancillotti}
\ead{emilio.ancillotti@iit.cnr.it}

\author[addr2]{Carlo~Vallati\corref{cor1}}
\ead{carlo.vallati@iet.unipi.it}

\author[addr1]{Raffaele~Bruno}
\ead{raffaele.bruno@iit.cnr.it}

\author[addr2]{Enzo~Mingozzi}
\ead{enzo.mingozzi@iet.unipi.it}

\cortext[cor1]{Corresponding author}

\address[addr1]{Institute for Informatics and Telematics (IIT) -- CNR, V. Giuseppe Moruzzi 1, 56124 Pisa, ITALY}
\address[addr2]{Dipartimento di Ingegneria dell'Informazione, University of Pisa, Via Diotisalvi, 2, 56122 Pisa, ITALY }

\thispagestyle{empty}

\begin{abstract}
Over the last few years, standardisation efforts are consolidating the role of the Routing Protocol for Low-Power and Lossy Networks (RPL) as the standard routing protocol for IPv6-based Wireless Sensor Networks (WSNs). Although many core functionalities are well defined, others are left implementation dependent.  Among them, the definition of an efficient link-quality estimation (LQE) strategy is of paramount importance, as it influences significantly both the quality of the selected network routes and nodes' energy consumption. In this paper, we present RL-Probe, a novel strategy for link quality monitoring in RPL, which accurately measures link quality with minimal overhead and energy waste. To achieve this goal, RL-Probe leverages both synchronous and asynchronous monitoring schemes to maintain up-to-date information on link quality and to promptly react to sudden topology changes, e.g. due to mobility. Our solution relies on a reinforcement learning model to drive the monitoring procedures in order to minimise the overhead caused by active probing operations. The performance of the proposed solution is assessed by means of simulations and real experiments. Results demonstrated that RL-Probe helps in effectively improving packet loss rates, allowing nodes to promptly react to link quality variations as well as to link failures due to node mobility. 
\end{abstract}


\maketitle

%
%
\section{Introduction}\label{sec:introduction}
\noindent
The future Internet of Things (IoT) foresees information systems seamlessly integrated with smart objects, i.e. daily objects empowered with computation and communication capabilities~\cite{2014_BORGIA,2015_KHOROV}. This vision will require sensors and actuators deployed on a large scale for ubiquitous sensing and remote control of physical systems. In this context, Wireless Sensor Networks (WSNs) will represent a key enabler to guarantee low-cost and rapid deployment of IoT devices exploiting multi-hop data delivery over wireless links. Efficiency and reliability of multi-hop data forwarding will be essential to support future IoT systems and guarantee their robustness \cite{stankovic14research}.

In general, the main objective of WSN routing protocols is to enable reliable communication while minimising resource consumption~\cite{zhao16energy}. This is particularly important in low-power and lossy networks (LLNs), because they are characterised by unreliable links whose quality may significantly fluctuate over time influenced by external interference or obstacles. In this context, the selection of the optimal path is however challenging, as gathering topology information requires communication and processing overhead that contrasts with the limited computational and energy capabilities available to constrained devices. 

Recently, significant efforts were put into defining a common routing protocol for IP-based WSNs, which have led to the standardisation of RPL, a gradient-based routing protocol that aims at building a robust multi-hop mesh topology over lossy links with minimal state requirements~\cite{rfc6550}. However, several studies have demonstrated that RPL is affected by reliability issues. The root cause of this unreliability is the lack of responsiveness to variations of network conditions that arise in real-life scenarios due to node mobility, or environmental factors, such as interference, multi-path effects, irregular radios patterns among the others~\cite{AncillottiBC12}. Hence, many extensions have been recently proposed for the original RPL specification to support routing optimisations and more efficient mechanisms for route discovery and topology repair, especially under mobility~\cite{Korbi2012_merpl,iscc14_mmrpl,ancillotti14reliable,Fotouhi2015_adhoc,momoro_rpl,Oliveira2016-comnet,rpl_iwcmc16,2017_ZHAO}. However, \emph{tweaking routing procedures does not solve the core problem of how to proactively maintain up-to-date information about links quality and network routes in a highly efficient manner}. 

More generally, the availability of a highly efficient, accurate, and adaptive link-quality estimation (LQE) framework is essential to allow any routing protocol to select the best routing path under time-varying network conditions. \added{It is well known that transmitting data over links with high quality improves both the network throughput, by limiting packet loss, and the network lifetime, by minimising the number of retransmissions. However, link quality estimation also plays a crucial role for detecting the dynamic behaviours of the links and maintaining the stability of the topology. For instance, LQE is needed to identify high quality links that are short-lived or to predict short-term variations of the quality of links. Then, costly route re-selection procedures that are triggered by link failures could be avoided}. Thus, LQE in wireless sensor networks has received significant attention over the past years~\cite{tang07channel,Baccour2012_ton}. Unfortunately, there are several limitations in using existing LQE techniques with RPL. In particular, broadcast-based probing is commonly adopted for LQE in low-power wireless networks because it incurs a lower overhead than unicast-based probing. However, RPL employs the Trickle algorithm~\cite{rfc6206} to dynamically control the dissemination of routing control information, which makes complicate the implementation of broadcast-based probing without affecting normal RPL operations. Furthermore, even with broadcast-based probing the measurement overhead increases almost linearly with the number of neighbours and the probing frequency, thus consuming precious energy resources and worsening network congestion. On the other hand, infrequent or on-demand probing would yield poor measurement accuracy and reduce routing ability to adapt to the link dynamics in real time. A few approaches have been proposed to support adaptive LQE in multi-hop wireless networks to minimise measurement overhead. Unfortunately, these existing techniques cannot be used in LLNs because they require the maintenance of large link-state tables or exploit cross-traffic overhearing~\cite{Kim2009_ton}. 

To address the above issues, in this paper we propose a novel \emph{lightweight link monitoring scheme}, called RL-Probe, which \added{is designed to ensure routing reliability (i.e., minimal packet loss rates) with lower overhead than classical periodic link probing, and without degrading the responsiveness to variable network conditions}. The salient features of RL-Probe can be summarised as follows. First, RL-Probe \emph{combines synchronous and asynchronous monitoring mechanisms} to maintain up-to-date information about link quality and their temporal variations while promptly reacting to sudden and unpredictable changes in network and link conditions. Secondly, RPL-probe uses \emph{a reinforcement learning technique based on the multi-armed bandit model}~\cite{mab} to dynamically control the probing procedures in order to automatically minimise measurement overhead without affecting responsiveness to route changes. Thirdly, RL-Probe preserves backward compatibility with standard RPL, i.e., enabling the interoperability of standard and enhanced nodes in the same network. It is also important to point out that the objective of RL-Probe is not to provide a novel routing metric for RPL, but a new measurement methodology that can be applied to generic link metrics, such as ETX~\cite{DeCouto2003}.

To evaluate the effectiveness of our solution, we have integrated RL-Probe within the Contiki RPL/6LoWPAN protocol stack. The proposed solution has been compared against the standard RPL configuration, in which no active probe is employed, and RPL with periodic probing, in which active probe traffic is exploited to monitor links and handle channel quality variations. In addition to static scenarios, experiments involving mobile nodes have been run to analyse the performance of RL-Probe in challenging conditions. Indeed, mobile nodes are characterised by rapidly-changing link quality conditions that are usually managed using ad-hoc extensions of RPL rather than standard LQE solutions. In order to offer a fair comparison, mRPL, a state-of-the-art RPL enhancement explicitly designed to cope with mobility, is also considered in the scenarios in which mobility is involved. We conduct both simulations and experiments in an indoor testbed in a broad range of static and dynamic scenarios. \added{We show that legacy RPL can suffer from packet loss rates of up to 65\% with mobile nodes, while RL-Probe achieves a packet loss rate lower than 15\% in the same scenarios. Furthermore, our results indicate that RL-Probe performs similarly to mRPL in terms of routing reliability, but it generates up to 30\% less control messages and consumes up to 20\% less energy per successfully transmitted packet. Finally, RL-Probe is faster than mRPL to detect sudden link quality variations thanks to the analysis of link trends, which allows RL-Probe to anticipate changes of link characteristics}.

The remaining of this paper is organised as follows. Section~\ref{sec:related} provides an overview of related work. In Section~\ref{sec:framework}, we present RL-Probe and we evaluate its performance in ~Section~\ref{sec:results}. Finally, in Section~\ref{sec:conclusion} we draw our conclusions.  
%
%
\section{Background and Related Work}\label{sec:related}
\noindent
In the following subsections, we first provide background information on RPL. Then, we overview the works that are the most relevant to our study. Specifically, we review both LQE approaches for RPL aimed at monitoring links quality, and mechanisms proposed to improve mobility support in RPL. We also discuss solutions that apply machine learning techniques to LQE.
%
%
\subsection{Background on RPL}\label{sec:rpl}
\noindent
%
RPL is an IPv6-based routing protocol specifically designed for lossy environments and resource-constrained embedded devices~\cite{rfc6550}. Specifically, RPL employs a distance vector routing algorithm which builds a logical topology on top of the physical network. In particular, the topology is a \emph{Destination Oriented Directed Acyclic Graph}, \emph{DODAG} for short. The root node of the DODAG initialises the DODAG formation by emitting DODAG \emph{Information Object} messages (hereafter \emph{DIOs} for short). Non-root nodes listen for DIOs and use the included information to join a DODAG. As a node joins a DODAG, it starts advertising its presence through the emission of DIO messages. Each DIO message specifies the rank of the sender, which is a scalar measure of the distance of that node from the root\footnote{The rank of each node is computed on the basis of an \emph{Objective Function} (\emph{OF} for short), which also defines how nodes select parents. Although the rank is not meant as a path cost, it is typically obtained from path metrics that are somehow function of the distance from the root, e.g., number of hops or each-to-end packet delays~\cite{2016_DIMARCO}.}. More recently, content-based approaches for path computation have bene also integrate in RPL~\cite{2016_AMADEO,2016_JIN}. Note that RPL can virtually split the network into multiple RPL Instances, which transport each kind of data according to its particular objective function~\cite{2016_BARCELO}. To avoid loops in the logical topology the rank must monotonically decrease along an upward path towards the DODAG root. 

As DIO messages are received from the neighbours, each node updates its view of the topology. In particular, node's neighbours with lower rank are selected to form a \emph{parent set} which is used for data forwarding. Among them, a \emph{preferred parent} is selected to forward traffic towards the root. Other important RPL control messages are the: (i) \emph{DODAG Information Solicitation} (\emph{DIS}), which is a multicast message used to trigger the transmission of DIO messages from neighbours; and (ii) \emph{Destination Advertisement Object} (\emph{DAO}), which is propagated upward (along the DODAG) to build the downward routes. To reduce the overhead associated to routing signalling RPL use a Trickle-based strategy~\cite{Levis04_trickle,rfc6206}, an adaptive beaconing scheme that exponentially increases the transmission timers when the network conditions are considered stable. \added{This implies that the Trickle communication rate is not periodic. As better explained later, this may negatively affect the ability of RPL to quickly detect topology changes using routing control packets}. Note that when a network inconsistency is detected (e.g., a network loop or preferred parent change) a repair procedure is triggered. This repair mechanism can be local (e.g., the node detaches from the DODAG and tries to reconnect) or global (the current DODAG is invalidated and a new DODAG is built). \added{Finally, Contiki also supports the Neighbour Discovery Protocol (NDP, defined in RFC 4861~\cite{rfc4861}), which implements four types of ICMPv6 messages for the purpose of router advertisements and neighbour unreachability detection. However, when RPL is activated the router advertisement procedures of NDP are disabled as DIO and DAO messages are already used to announce the presence of other nodes and their link parameters. Furthermore, when receiving DIO or DAO packets, RPL populates the NDP neighbour info table. Then Neighbour Solicitation (NS) and Neighbour Advertisement (NA) messages are generated by the Neighbour Unreachability Detection (NUD) mechanism that is included in NDP to refresh the table and to verify that a neighbour is still reachable through a cached link layer address}.
%
%
%
%
%
\subsection{LQE in RPL}\label{sec:lqe_rpl}
\noindent
There is a large body of research on LQE in WSNs and it is out of the scope of this paper to overview the several link quality metrics that exist in the literature (the interested reader is referred to the comprehensive survey in~\cite{Baccour2012_ton}). 

From a general perspective, an LQE framework consists of three components: $i)$ \emph{link monitoring}, which is the mechanism used to collect link measurements; $ii)$ \emph{link measurement}, which specifies the information to be retrieved, and $iii)$ \emph{metric evaluation}, which defines the metric to assess the link quality. First RPL implementations employed simple passive monitoring techniques, which leverage statistics of transmission failures for the links used by data traffic~\cite{DawansDB12}. However, data-driven link estimation methods do not apply to idle links, and the reactivity of such methods heavily depends on the network traffic patterns. Furthermore, overhearing is required to monitor links to neighbouring nodes that are not the preferred parent~\cite{zhang10comparison}. Thus, recent RPL implementations prefer active monitoring to measure link quality through probes. The Contiki 3.0 RPL implementation, for instance, includes an optional probing mechanisms (dubbed RPL-PP), in which neighbours are probed periodically through unicast DIO messages. The target of the probe is selected following a round robin strategy, except the link to the preferred parent that pre-empts the other links to neighbouring nodes in the probing schedule if its quality has not been updated in a given interval.

It is generally agreed that unicast-based probing provides accurate link measurements~\cite{zhang09link}. The downside is that neighbours are assessed individually, which results into long convergence times for link quality estimation, especially when large or dense networks are considered. In order to reduce the measurement delay, a broadcast-based probing has been proposed in many RPL extensions. This is implemented in different ways, e.g. forcing the periodicity of routing control messages~\cite{co-rpl}, or sending bursts of RPL control messages during specific phases of network operations (e.g. at network formation) or at the occurrence of certain events (e.g. during parent changes)~\cite{Fotouhi2015_adhoc}. Those schemes are sender-initiated, as the transmitting node generates the probe packets. One the other hand, some RPL variants use receiver-side probing schemes, which use information from received probes to estimate link quality. For instance, in~\cite{ancillotti14reliable} each node triggers the link monitoring by sending a multicast DIS message to its neighbours, which reply with a train of unicast DIS messages. The advantage is that multiple neighbours can be monitored in parallel even if unicast probes are used. One potential drawback of this approach is that the metric computation is performed on the opposite direction to data transmission. However, many experimental studies show the symmetry of wireless links in real use cases of WSNs~\cite{tang07channel,Srinivasan2010}. 

All the above-described mechanisms adopt basic measurement schemes. However, there are a few examples of more sophisticated solutions for LQE, which use a hybrid approach by combining multiple complementary methods. For instance, a link-quality measurement framework is proposed in~\cite{Kim2009_ton}, which combines three measurement techniques: passive, cooperative, and active monitoring. However, this framework cannot be easily applied in low-power wireless networks because it requires the maintenance of large link-state tables and leverage cross-traffic overhearing. Differently from this prior work, RL-Probe proposes a lightweight framework that can be implemented in constrained devices in which different methods for LQE are adopted.

%
%
\subsection{Machine Learning for LQE}\label{sec:lqe_learning}
\noindent
\added{During the past decade, machine learning algorithms and computation intelligence methods have been increasingly applied in wireless sensor networks to improve network performance and solve a variety of networking problems~\cite{Alsheikh2014learning,Kulkarni11_ai}. In particular, prediction models that utilise different machine learning algorithms have been frequently proposed to automatically estimate link quality. For instance, a logistic regression classifier is presented in~\cite{tosn14_Liu2} that uses a combination of physical parameters, such as RSSI and SNR, to predict the success probability of the next transmission. An online learning algorithm based on artificial neural networks is described in~\cite{tosn14_Liu1} to predict short-term quality variations. Decision trees and classification rules for supervised learning of link quality are developed in~\cite{metricmap}. Online channel quality estimation is modelled as a game of prediction with expert advice in~\cite{Marinca15learning}. Unsupervised feature selection is proposed in~\cite{Panousopoulou16learning} to determine the dominant features for the performance of end-to-end links. Multi-arm bandit problems have been also proposed to deal with the channel exploration-exploitation dilemma in wireless networks. For instance, the authors in~\cite{Toldov16learning} formulate the channel selection problem in cognitive radio networks using a multi-arm bandit model, and propose a fast converging sampling method. The channel decision problem is modelled as a restless multi-arm bandit problem in~\cite{Wang15MAB}. In the context of computational intelligent algorithms for link quality prediction, fuzzy logic is frequently used. For instance, authors in~\cite{momoro_rpl} and~\cite{Baccour10_ewn} propose fuzzy rules to combine multiple link metrics while compensating for the uncertainties in the wireless channel conditions.}

\added{Our work differs from the above-mentioned studies, as we use machine learning techniques to increase the efficiency of link sampling and not to develop new predictive models of link performance. Multi-armed bandit problems have been extensively applied in the context of channel selection in the cognitive radio context. However, the channel probing problem is different from the link measurement problem considered in this paper, as in the former case the objective is to infer the binary channel state (unoccupied or busy), while in this study we focus on optimising the trade-off between probing frequency and responsiveness in detecting short-term link quality variations}.
%
%
%
\subsection{RPL Mobility Extensions}\label{sec:related_mobility}
\noindent
A survey of recent RPL extensions and modifications to improve mobility support is provided in~\cite{Oliveira2016-comnet}. Previous studies have shown that RPL provides a fast network setup but that mobility support is not adequate and it should be improved~\cite{AccenturaGBC11}. A few works have investigated neighbour discovery protocols that improve the capability of mobile devices to remain connected to the WSN as they move~\cite{2016_WU,2017_MONTERO}. However, most of existing solutions for mobility management in RPL focus on modifying native RPL mechanisms to improve RPL ability of detecting link failures, or to speed up the handoff and local repair procedures~\cite{2016_BOUAZIZ}.

In ME-RPL~\cite{Korbi2012_merpl} nodes are separated as \emph{mobile} nodes (i.e. nodes with unstable links) and \emph{fixed} nodes, with mobile nodes forced to act only as leaf nodes to avoid network paths through them. Then, mobile nodes must advertise their mobility status in control messages and generate frequent DIS messages to update their parent set information. The time between DIS messages is computed based on the frequency of preferred parent changes using a simple multiplicative increase/multiplicative decrease algorithm.  

MoMoRo~\cite{momoro_rpl} quickly reacts to packet transmission failures by immediately resending the lost packet, and starting a new route search (by broadcasting beacons and collecting updated routing information from the neighbours) if the second attempt also fails. Furthermore, MoMoRo uses a fuzzy estimator that combines different link metrics (ETX, average RSS, symbol error rate variance) to classify links based on their probability of disconnection. 

As for ME-RPL, the authors in~\cite{iscc14_mmrpl} proposes that mobile nodes only connect as leaves in the DODAG. Furthermore, a reverse Trickle timer -- the timer starts from the maximum value and halves the DIO sending intervals after each new DIO transmission -- is used by the preferred parents of mobile nodes to trigger DIO messages. The implicit assumption of this approach is that mobile nodes' stability decreases with time. An adaptation of the Trickle timer algorithm for better mobility support is also proposed in mod-RPL~\cite{rpl_iwcmc16}. Specifically, mod-RPL takes into account the trajectory and velocity of mobile nodes when selecting the sending interval of DIO messages, and it dynamically adapts the timer to the distance between the mobile node and its preferred parent.

KP-RPL is a position-based extension of RPL to provide mobility support~\cite{Barcelo2016_sensor}. Standard RPL is used for routing between fixed nodes, while mobile nodes make their routing decisions using positioning information obtained by applying a Kalman filter on RSSI measurements. Furthermore, mobile nodes generate a blacklist to discard neighbours that could not be reachable due to positioning errors.

In conclusion, all the above-described schemes are specifically designed to handle mobility through ad-hoc mechanisms that usually do not perform accurate LQE. Differently from such solutions, our proposed approach is not restricted to support only node mobility, but it can seamlessly manage node mobility through fine-grained LQE, without requiring modifications of standard RPL or a-priori knowledge of which nodes are mobile.

%
\paragraph{Overview of mRPL}\label{sec:mRPL}
\noindent
We present mRPL~\cite{Fotouhi2015_adhoc} in more details as it will be used as benchmark in the following evaluation\footnote{Note that a recent enhancement of mRPL, called mRPL+, has been presented in~\cite{2017_FOTOUHI}. The core mechanisms of mRPL and mRPL+ are similar and our findings can be applied to mRPL+ as well.}. 

Basically, mRPL integrates smart-Hop, a handoff scheme for low-power networks~\cite{Fotouhi2014_tmc}, in RPL. As in~\cite{Korbi2012_merpl,iscc14_mmrpl,Barcelo2016_sensor} mobile and fixed nodes are separated. Then, mobile nodes monitor the link quality by receiving DIO messages from their preferred parents. A mobile node disconnects from the preferred parent and enters a discovery phase if the average received signal strength is below a given threshold or if no packets are received by the parent before a connectivity timer ($T_C$) expiration. To avoid that the high variability of wireless links causes frequent handoffs a hysteresis mechanism is applied to this RSSI threshold. During the discovery phase, the mobile node multicasts DIS messages to all neighbouring nodes and collects their unicast DIO replies to decide which new preferred parent to select based on the average RSSI level. As an additional stability mechanism, a mobile node repeats the discovery procedure $m$ times after switching to a different preferred parent to check the stability of that node. 

mRPL introduces additional timers to increase handoff efficiency and reliability. In particular, the mobility detection timer ($T_{MD}$) is used to detect connection losses due to the existence of external objects (obstacles between the sender and the receiver). The handoff timer establishes the transmission period of DIS messages to the neighbouring parents. Finally, a reply timer ($T_R$) is used to select the time instant at which a parent should reply to the mobile node to reduce the probability of collision between control messages during the handoff process. 

\added{It is important to emphasise that mRPL is not as general as RL-Probe, since it requires to a priori configure nodes as either mobile or static (called APs) in order to perform different operations. In fact, mobile nodes are restricted to be leaf nodes in the RPL DODAG. This may cause inefficient routing in fully static networks, as also shown in Section~\ref{sec:results}.}
%
%
%
%
\section{The RL-Probe framework}\label{sec:framework}
\noindent
This section describes RL-Probe in details. First, we provide background material on the multi-armed bandit model. Then, we explain the design rationale behind RL-Probe. Finally, we detail the main mechanisms and algorithms used in RL-Probe. 
%
%
%
%
\subsection{Multi-armed Bandit Background}\label{sec:MAB}
\noindent
Generally, a multi-armed bandit (MAB) model is used to describe a learning problem in which an agent must repeatedly choose among different options, or actions~\cite{mab}. After each choice, the agent receives a reward chosen from an \emph{unknown} stationary probability distribution that depends on the selected action. The objective of the agent is to maximise the expected total reward over a time horizon. The MAB model is so named by analogy to a slot machine that has multiple levers, with different but unknown probabilities of hitting the jackpot. Then, the player should try to find the best levers that maximise the winnings by repeatedly playing. 

More formally, let us assume that time is discretised and time slots are numbered as $n=1,2,\ldots$. Then, let $\mathcal{U}$ be the set of actions (or arms). At round $n$, the arm pulled is $u^{(n)}$ and the reward received is $W^{(n+1)}$. We assume that the reward provided by arm $u$ follows a random distribution $F_{u}(x)$, which is unknown. Now, let $s^{(n)}$ be the representation of the system state at round $n$ and let $\mathcal{S}$ be the discrete set of possible states. The history of the system at a given stage is the sequence of decisions, observed states and collected rewards. For the sake of tractability, MAB models usually assume the \emph{Markov property}, i.e., rewards depend only on the current state and the current action and not on the full history of previous actions and states. Then, the core of a MAB model is the policy $\pi$, namely the mapping function between states and actions, which should maximise the amount of rewards the agent receives over time. Several methods have been proposed in the literature to learn the optimal policy without requiring a model of the system behaviours, but leveraging only on the experience obtained by iteratively interacting with the system. As discussed more in depth in the following sections, these learning techniques have to cope with the \emph{exploration/exploitation dilemma}, which implies balancing immediate gains (i.e., selecting the action with the maximum expected reward) with knowledge creation to make better decisions in the future (i.e., selecting actions that appear to be worse but could potentially be the best). Typically, a probabilistic learning strategy is defined that assigns a probability to each possible action in a state according to an estimation of the current state value. 
%
%
%
\subsection{Overview of RL-Probe}\label{sec:overview}
\noindent
One of the main distinct features of RL-Probe over existing probing strategies for LLNs is that it adopts a \emph{hybrid} approach to adaptively combine \emph{synchronous} and \emph{asynchronous} LQE techniques. More precisely, the synchronous LQE technique relies on unicast probes to provide accurate link quality measurements. However, we design two novel mechanisms to make unicast-based probing more adaptive and responsive, without introducing excessive probing overhead. First, we formulate the selection of the probing period as a multi-armed bandit problem to dynamically adjust the probing frequency to the link variability in real time. Our learning-based approach provides a good trade-off between overhead and responsiveness to varying link quality. Secondly, we cluster neighbours into groups and we assign different probing priorities to each group. The clustering and the priority selection are based on the importance of each node in RPL route maintenance and recovery procedures. The rationale behind this approach is that wireless link correlation (e.g., due to cross-network interference under shared medium) has been observed in many recent studies~\cite{twc16_correlation}. Consequently, independent estimates of \added{individual link quality can lead to redundant measurements}, especially in dense networks. 
\begin{algorithm}[tp]
		\begin{algorithmic}[1] 
		\small
			\Require{$\mathcal{N}_{i}$}\Comment{set of neighbours}
			\Loop
			\If{(received packet $p$ from $j$)}
				\State Get RSSI value ($rssi$) from packet $p$; \label{alg:rssi}
				\State $rssi_{i,j}[].add(rssi)$; \label{alg:rssi_list}
				\State Get RSSI trend ($rssiTrend_{i,j}$) for link $l_{i,j}$ from the last $n$ RSSI samples; \label{alg:rssi_trend}
				\If{$(p = \text{ACK}$ \textbf{and} $j=pp)$}
					\State $\Delta rssi_{i,j} = \frac{rcvTh- rssi }{rcvTh}$;
					\If{$(rssiTrend_{i,j} <0$ \textbf{and}  $\Delta rssi_{i,j} \le \alpha)$} \label{alg:proactive}
						\State \textbf{do} receiver-side probing; \label{alg:multi1}
						\ForAll{$k \in \mathcal{N}_i$}
							\State update link stability ($cv_{i,k}$) for $l_{i,k}$; \label{alg:uLQE1}
						\EndFor
					\EndIf
				\ElsIf{$(p = \text{NACK}$ \textbf{and} $j=pp$ \textbf{and} $cv_{i,j}[].get(last)\le \beta)$} \label{alg:reactive}
					\State \textbf{do} receiver-side probing; \label{alg:multi2}
					\ForAll{$k \in \mathcal{N}_i$}
						\State update link stability ($cv_{i,k}$) for $l_{i,k}$; \label{alg:uLQE2}
					\EndFor
				\EndIf
			\ElsIf{(timer $T_{p}$ expires)}
				\State update nodes in set $\mathcal{P}_i$ and $\mathcal{O}_i$; \label{alg:updatecluster}
				\State $x \gets Uniform(0,1)$;
				\If{$(x \le \epsilon)$}\Comment{exploitation phase}\label{alg:exploitation1}
					\State $u \gets \underset{x}{\operatorname{argmax}}\left ( W^{(x)}[].get(last)\right )$; \label{alg:exploitation2}
				\Else \Comment{exploration phase} \label{alg:exploration1}
					\State $u \gets random(D_1, D_2, D_3)$; \label{alg:exploration2}
				\EndIf
				\If{$(x=D_1)$}
					\State  $j \gets \textsc{BestNode}(\mathcal{P}_i)$
				\ElsIf{$(x=D_2)$}
					\State $j \gets \textsc{BestNode}(\mathcal{O}_i)$
				\EndIf
				\If{$(x \ne D_3)$}
					\State \textbf{do} unicast probing to $j$;
					\State update link stability ($cv_{i,j}$) for $l_{i,j}$;
					\State \textsc{UpdateUtility}($i,j$);
				\EndIf
				\State \textsc{UdapteReward}(i,j,u)
			\EndIf
			\If{(sent data packet to $pp$)}
				\State update link stability ($cv_{i,j}$) for $l_{i,j}$;
				\State \textsc{UpdateUtility}($i,pp$); \label{alg:utility_pp}
			\EndIf
			\EndLoop
		\end{algorithmic}
\caption{Main procedure of RL-Probe}\label{alg:rl-probe}
\end{algorithm}

Our asynchronous LQE mechanism is designed to efficiently handle sudden and disruptive link variations. To this end, we integrate in RL-Probe a receiver-side probing method first proposed in~\cite{ancillotti14trickle}, which allows to rapidly assess the quality of the links from a node to all its neighbours with a sufficient accuracy. In principle, asynchronous probing could facilitate the isolation of faulty nodes/links, or the detection of preferred parent unavailability due to mobility. Clearly, asynchronous probing must be activated on-demand when it is most likely needed because it is costly in terms of energy and bandwidth consumption. Thus, we also propose specific triggering mechanisms for our asynchronous LQE technique, which are based on the trend of received signal strength indicator (RSSI) and link quality \added{(ETX)} values. Our solution reduces the cost of recovering RPL connectivity by ensuring quick detection of network disruptions without depleting the limited resources of the devices. \added{Finally, it is worth pointing out that RL-Probe is specifically designed for single-channel MAC protocols. A number of multi-channel MAC protocols for low-power sensor networks, which perform channel hopping rather than allocating fixed channels to data collection trees, have been also studied~\cite{MiCMAC}. However, how to implement efficient broadcast and unicast probing devices switches periodically between channels is still an open research issue.}

For the sake of clarity, the pseudocode of the main mechanisms of RL-Probe are provided in Algorithm~\ref{alg:rl-probe}, Algorithm~\ref{alg:rl-function-utility}, Algorithm~\ref{alg:rl-function-reward} and Algorithm~\ref{alg:rl-function-bestnode}. 
%
%
%
\subsection{Asynchronous Probing}\label{sec:async_probing}
\noindent
As outlined above, the proposed asynchronous probing scheme exploits the measurements of RSSI values \added{and ETX metric} to detect sudden link variations. More precisely, whenever a node $i$ receives a packet from a neighbour $j$, it obtains the RSSI value from the wireless transceiver (line~\ref{alg:rssi} in Algorithm~\ref{alg:rl-probe}). A list of \added{the most} recent RSSI values is maintained for each link $l_{i,j}$ (line~\ref{alg:rssi_list} in Algorithm~\ref{alg:rl-probe}), which is used to estimate the trend in RSSI variations (line~\ref{alg:rssi_trend} in Algorithm~\ref{alg:rl-probe}). For the sake of computational efficiency, in our implementation we estimate the RSSI trend using a simple moving average (SMA) filter over the last \added{four} measurements of RSSI differences between consecutively received packets. \added{More formally, each node $i$ maintains a list with the previous $n$ RSSI samples it has received from neighbour $j$. If $rssi_{i,j}[k]$ denotes the last received RSSI sample, this list consists of the values $(rssi_{i,j}[k], rssi_{i,j}[k-1],\ldots,rssi_{i,j}[k-(n-1)])$. Then, the RSSI trend is computed as the unweighted mean of the previous three RSSI differences. This can be written as}:
\begin{equation}
rssiTrend_{i,j}[k] = \sum_{l=0}^{n-1}  \frac{\left(rssi_{i,j}[k] - (rssi_{i,j}[k-l]\right ) }{n} \; ,
\end{equation}
\added{with $n=4$ in our implementation. A similar SMA filter is also maintained for the ETX measurements obtained from data packets and unicast probes.} Then, our asynchronous probing has both a \emph{proactive} and \emph{reactive} phase. The proactive phase tries to anticipate topology changes and it is activated when a packet is successfully transmitted to the preferred parent $pp$\footnote{As introduced in Section~\ref{sec:rpl}, the preferred parent of a node on a path towards the  the DODAG root is the parent with the lowest rank value.} (i.e., the sender receives a MAC ACK from the preferred parent) \added{but a set of simultaneous conditions occur that suggests a possible degradation of the link quality. Specifically, the following three conditions must be satisfied to trigger a receiver-side probing}\footnote{We recall that during a receiver-side probing phase a node sends a multicast DIS message and its neighbouring nodes reply with a train of unicast DIS messages.} (lines~\ref{alg:multi1}-\ref{alg:uLQE1} in Algorithm~\ref{alg:rl-probe})\added{: $i)$ the RSSI trend is negative; $ii)$ the last RSSI sample is close to the receiver sensitivity $rcvTh$, and $iii)$ the ETX trend is negative}. 

On the other hand, the reactive phase is designed to facilitate local repair operations after unexpected network disruptions. In this case, the receiver-side probing is activated when there is a transmission failure (i.e., the sender receives a MAC NACK from the preferred parent) on a link that has a stable link quality. The link quality stability is measured using the coefficient of variation of the link quality (defined as the ratio between the standard deviation $\sigma_{i,j}$ and the mean $\mu_{i,j}$). More formally, a link is considered stable if $cv_{i,j} \le \beta$ (line~\ref{alg:reactive} in Algorithm~\ref{alg:rl-probe}). Note that the selection of threshold $\beta$ is dictated by the variability of the wireless links. Links with $cv_{i,j}$ lower than one are considered low-variance. Summarising, both a successful transmission on a rapidly degrading link and a packet loss on a stable link activates the asynchronous probing for updating the link quality to all nodes in the neighbourhood (lines~\ref{alg:multi2}-\ref{alg:uLQE2} in Algorithm~\ref{alg:rl-probe}). 
%
%
%
%
\subsection{Synchronous Probing}\label{sec:sync_probing}
\noindent
Differently from conventional unicast-based probing schemes, our synchronous probing does not use a fixed probing interval, which would cause unacceptable convergence delays for the link quality estimation, especially in dense networks. On the contrary, RL-Probe clusters nodes into separate groups and adaptively adjusts the probing period for each group. Different approaches for such link clustering can be devised, taking also advantage of cross-layer information from the network layer. For instance, in this study we define a link clustering that considers the importance of a neighbour to maintain good RPL connectivity. More precisely, let us denote with $\mathcal{N}_{i}$ the set of neighbours of node $i$. We define the set $\mathcal{P}_{i}$ that contains the best $m_p$ parents of node $i$, i.e. parents that have the lowest path cost to the sink if selected as next hop by node $i$. Similarly, we define the set $\mathcal{O}_{i}$ that contains up to $m_o$ nodes from the set $\mathcal{N}_i \setminus \mathcal{P}_i$, which have the lowest path cost to the sink if selected as next hop by node $i$. The $m_p$ and $m_o$ parameters should be selected as a trade-off between reliability and responsiveness. Large $m_p$ and $m_o$ values provide coarse grained information about the links and decrease the responsiveness of the system. On the other hand, low $m_p$ and $m_o$ values reduce the possibility to discover good alternative paths in case of loss of the preferred parent. In general, $m_o > m_p$ as it is more critical to have detailed information on nodes of the parent set. 

The decision-making process that determines which set to probe and how frequently to probe it is formalised through a MAB model. Specifically, we define three possible actions as follows:  
\begin{itemize}
\item[$D_1$]: probe a node in $\mathcal{P}_{i}$;
\item[$D_2$]: probe a node in $\mathcal{O}_{i}$;
\item[$D_3$]: skip the probing.
\end{itemize}
Each of the above decisions corresponds to an independent arm in the MAB problem. The reward associated to each probing decision, say $W^{(x)}$ with $x \!\in\! \{D_1,D_2,D_3\}$, corresponds to the potential \emph{gain} provided by probing a node in a specific group. For instance, the gain can be a measure of the improved network responsiveness to link quality variations. Details on reward estimation are provided later in this section. 

First of all, let us explain which is the policy used by node $i$ to select the probing action given the knowledge of the average rewards. In this study, we adopt the well-known $\epsilon$\emph{-greedy} algorithm, which selects with probability $\epsilon$ the action with the maximum accumulated reward, or \emph{greedy action} (lines~\ref{alg:exploitation1}-\ref{alg:exploitation2} in Algorithm~\ref{alg:rl-probe}), and with probability $(1-\epsilon)$ selects a random action (lines~\ref{alg:exploration1}-\ref{alg:exploration2} in Algorithm~\ref{alg:rl-probe}). By properly tuning the $\epsilon$ parameter it is possible to balance exploration and exploitation phases to fast converge to the optimal action selection. If action $D_1$ or $D_2$ are selected it is necessary to decide which neighbour to probe in the set $\mathcal{P}_i$ and $\mathcal{O}_i$, respectively. As anticipated above, we exploit a measure of the utility of a node for the RPL topology maintenance procedure. Note that this utility measure is also used in the reward computation. In the following we formalise the algorithms used in RL-Probe to compute the utility and reward metrics.
\begin{algorithm}[tp]
		\begin{algorithmic}[1] 
		\small
			\Statex
			\Function{UpdateUtility}{$i,j$}\Comment{$l_{i,j}$ probed link} 
				\State $\omega_{i,j}[].add\left (\mu_{i,j}[].get(last) \!+\! \sigma_{i,j}[].get(last) \right)$; \label{alg:omega}
				\State $\Delta \omega_{i,j}[].add\left (\omega_{i,j}[].get(last) - \omega_{i,j}[].get(last-1) \right )$; \label{alg:delta_omega}
				\If {$(\Delta \omega_{i,j}.get(last) \cdot \Delta \omega_{i,j}.get(last-1) > 0 )$}  \label{alg:upos1}
					\State $U_{i,j} \gets U_{i,j} + \lvert \Delta \omega_{i,j}[].get(last) \rvert $;  \label{alg:upos2}
				\Else
					\State $U_{i,j}(n) \gets 0 $ ; \label{alg:unull}
				\EndIf
			\EndFunction
		\end{algorithmic}
\caption{Description of the function for utility update}\label{alg:rl-function-utility}
\end{algorithm}

\begin{algorithm}[tp]
		\begin{algorithmic}[1] 
		\small
			\Statex
			\Function{UpdateReward}{$i,j,u$} \Comment{$u$ MAB action} 
				\If {($u=D_1 )$} 
					\State $ W^{(D_1)}_{i}[].add \left (\max \left [ 0, \max\limits_{j \in P_i} U_{i,j}(n) \!-\! C_{1} \right ] \right )$ ; \label{alg:w1}
				\ElsIf {$(u=D_2 )$} 
					\State $ W^{(D_2)}_{i}[].add \left (\max \left [ 0, \max\limits_{j \in O_i} U_{i,j}(n) \!-\! C_{2} \right ] \right )$; \label{alg:w2}
				\Else
					\State $ W^{(D_3)}_{i}[].add \left (\max \left [ 0, G_{np} - U_{i,pp}(n) \right ] \right )$; \label{alg:w3}
				\EndIf
			\EndFunction
		\end{algorithmic}
\caption{Description of the function for reward update.}\label{alg:rl-function-reward}
\end{algorithm}

\begin{algorithm}[tp]
		\begin{algorithmic}[1] 
		\small
			\Statex
			\Function{BestNode}{$\mathcal{A}$} \Comment{candidate node to be probed} 
				\State $x \gets Uniform(0,1)$;
				\If{$(x \le \epsilon)$}\Comment{exploitation phase}
					\State Get node $i \in \mathcal{A}$ with the highest utility; \label{alg:bn1}
				\Else
					\State Get node $i \in \mathcal{A}$ randomly; \label{alg:bn2}
				\EndIf
			\EndFunction
		\end{algorithmic}
\caption{Description of the function for the selection of the best node.}\label{alg:rl-function-bestnode}
\end{algorithm}
%
%
%
\paragraph{Utility and reward computation} In RL-Probe, the reward function is mainly used to estimate the trends in link quality variations (e.g., quality degradation for an interfered link). To this end, we follow the same approach as in~\cite{jsac16_routing_metric} and we use the \emph{mean} and the \emph{standard deviation} of the link quality metric. More formally, let us assume that each node maintains a list of the estimated values of the mean $\mu_{i,j}$ and standard deviation $\sigma_{i,j}$ of link $l_{i,j}$\footnote{An exponential moving average (EMA) filter with smoothing factor 0.8 is used to update these estimates.}. We introduce an aggregate measure $\omega_{i,j}$ of the link quality variability, as the sum of $\mu_{i,j}$ and $\sigma_{i,j}$ (line~\ref{alg:omega} in Algorithm~\ref{alg:rl-function-utility}).We can easily compute the incremental variation of the link quality variability as the difference of two consecutive samples of $\omega_{,j}$ (line~\ref{alg:delta_omega} in Algorithm~\ref{alg:rl-function-utility}). Intuitively, it might be appropriate to monitor more frequently links that are showing a clear \emph{trend}, in order to timely identify a link that is quickly degrading (e.g., due to an external interference) or improving. Thus, we associate a \added{positive} utility to links that \added{showed the same trend of} link quality variation in the last two probes (lines~\ref{alg:upos1}-\ref{alg:upos2} in Algorithm~\ref{alg:rl-function-utility}), while we assign a null utility to links that are not characterised by a steady (positive or negative) trend (line~\ref{alg:unull} in Algorithm~\ref{alg:rl-function-utility}). \added{Clearly more sophisticated utility functions can be designed to utilise a large number of previous samples and more complex decision rules. However, our main objective is to determine the feasibility of our approach and we only check the trend of the last two samples for the sake of implementation simplicity}. Now, we can also clarify how the \textsc{BestNode}($\mathcal{A}$) function chooses the neighbour to probe in set $\mathcal{A}$. In the simplest case, it could select the node with the highest utility. However, this would make impossible to check, even infrequently, links with small utilities (i.e., more variable links). Following the same line of reasoning of the above-discussed $\epsilon$-greedy exploration strategy, the \textsc{BestNode}($\mathcal{A}$) function selects the node with the highest utility with probability $\epsilon$ (line~\ref{alg:bn1} in Algorithm~\ref{alg:rl-function-bestnode}), while a random link in the set $\mathcal{A}$ in the other cases (line~\ref{alg:bn2} in Algorithm~\ref{alg:rl-function-bestnode}). 

Commonly, reward functions for learning problems should include a positive term and a negative term to be well specified. The positive term measures the gain of performing that action. For the case of actions $D_1$ and $D_2$ the gain is given by the highest utility value in the group (lines~\ref{alg:w1} and~\ref{alg:w2} in Algorithm~\ref{alg:rl-function-reward}). Thus, the reward of a link cluster is high if there is at least one link in the set with a consistent variability pattern for its link quality. Intuitively, the cost for actions $D_1$ and $D_2$ should be a measure of the cost of a unicast probe. Since we want to give higher priorities to probing parents than other neighbours, we have that $C_1 \!\ge\! C_2$. For the same reason, we assume that skipping a probing phase provides a gain $G_{np}$, in terms of saved node and network resources (line~\ref{alg:w3} in Algorithm~\ref{alg:rl-function-reward}). As a cost for the action $D_3$ we use the utility of the link with the preferred parent because if the link with the preferred parent is not stable a node should keep looking for alternative links. 
%
%
%
%
\paragraph{Group management} As explained above link clustering is controlled using the path cost. It is important to point out that link quality variations, especially for neighbours with intermediate link quality, may yield to frequent changes in the cluster composition. However, this can negatively affect the convergence of the learning algorithm. Therefore, we define a \emph{hysteresis margin} for the sojourn time of a node in the set $\mathcal{P}_i$. Specifically, when a node $j$ in the set $\mathcal{P}_{i}$ is not anymore among the best $m_p$ neighbours for node $i$ it would be removed only if this condition persists for at least a time $t_{hyst}$. This check is implemented in the function \textsc{UpdateClusters}($\mathcal{P}_i,\mathcal{O}_i$) (line~\ref{alg:updatecluster} of Algorithm~\ref{alg:rl-probe}). 
%
%
%
%
\paragraph{Preferred parent monitoring} In RL-Probe the link quality to the preferred parent is estimated by passively monitoring the data traffic, as in legacy RPL. For this reason, the preferred parent is not part of the set $\mathcal{P}_i$. However, the link quality measurements are still used to update the utility estimates for the preferred parent (line~\ref{alg:utility_pp} of Algorithm~\ref{alg:rl-probe}).

%
%
\section{Performance Evaluation}\label{sec:results}
\noindent
In order to implement and evaluate RL-Probe, we opted for the Contiki 3.0 operating system (OS). The main reasons for selecting Contiki are: $i)$ the support of Cooja simulator, which allows to easily port the software on real hardware, (ii) the availability of a standard RPL implementation that is widely used, and (iii) the availability of several plugins that already implement mobility models, interference models and probing techniques. Table~\ref{tab:parameters} summarises the RL-Probe parameters used in the algorithms described in Section~\ref{sec:framework}, which have been fine-tuned with extensive simulations.

In this section, we evaluate RL-Probe against standard implementations of legacy RPL, RPL-PP and mRPL using both simulations and experiments. Specifically, we consider a basic RPL implementation that measures the quality of links through passive monitoring of the links used by data traffic~\cite{DawansDB12}. Secondly, we consider RPL-PP, included in Contiki 3.0 RPL implementation, as a solution specifically designed to handle link quality variations. Finally, mRPL is also considered as a term of comparison, to verify the efficacy of RL-Probe in handling mobility. To this aim, we ported the mRPL implementation described in~\cite{Fotouhi2015_adhoc} and available for Contiki 2.6.1 to the latest version of the Contiki OS. It is important to recall that mRPL needs to differentiate between mobile nodes and fixed nodes (called Access Points), and mobile nodes are forced to act as leaf nodes in the routing tree. For this reason, we compare mRPL with RL-Probe only in scenarios in which mobility is involved, or there are many unstable links.

\paragraph{Metrics} We consider three main performance metrics in our evaluation. First, we measure the packet loss ratio (PLR) at the sink, defined as the percentage of failed packet transmissions over the total number of packets sent by a node. Secondly, we consider the packet overhead measured as the sum of the RPL control messages, including probe packets. Thirdly, we measure the \added{normalised energy consumption per successfully received packet at the sink. We argue that this ensures a fairer comparison between scenarios affected by different PLR values than considering the total energy consumed per unit of time}. 
\begin{table}
\caption{RL-Probe Protocol Parameters.\label{tab:parameters}}
\centering 
\begin{tabular}{| l | c |}
\hline
Parameter & value \\
\hline\hline
 $\alpha$&  3\%\\
 $\beta$&  1\\
 $m_p/m_o$ & 3 / 10\\
$C_{1}/ C_{2}$ & 1 / 5\\
 $G_{np}$ & 10 \\
 $\epsilon$ & 0.7 \\
 $t_{hyst}$ & 10~minutes\\
 $T_{p}$ & 1 minute \\
\hline
\end{tabular}
\vspace{-0.2cm}
\end{table}

%
%
%
\subsection{Simulation Analysis\label{sec:sim_analysis}}
\begin{figure}[ht] 
\centering
\includegraphics[trim={2cm 7cm 1cm 6.5cm},clip,angle=0,width=0.5\textwidth]{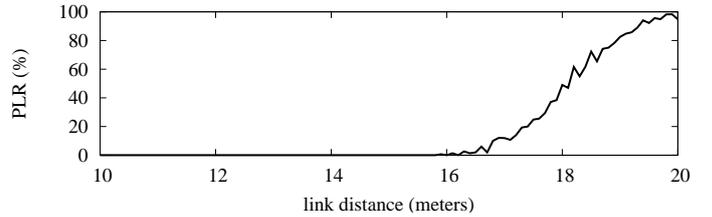}
\caption{Link characterisation in Cooja simulator under the MRM model.\label{fig:mrm}}
\vspace{-0.3cm}
\end{figure}

\noindent
A simulation study is needed to investigate the performance in controllable and easily reproducible network conditions. Thus, we use Cooja to simulate Tmote Sky nodes. In WSNs, a radio duty cycling (RDC) mechanism is typically implemented to switch on and off the radio transceiver in order to save energy. ContikiMAC is the default RDC scheme used in our tests~\cite{contikimac}. To model realistic radio propagation and interference we use the Multipath Ray-tracer Medium (MRM), which supports multi-path effects~\cite{Osterlind06_mrm}. We have configured MRM parameters to achieve a 100\% success rate at 10~meters and an interference range of 20~meters. Figure~\ref{fig:mrm} shows the packet loss rate as a function of node distance for a link under the MRM model. The traffic flows generate a Constant Bit-Rate (CBR) traffic consisting of small UDP messages (40 bytes) sent every one minute from all the nodes to the sink. \added{We select a CBR traffic model as it well represents the period traffic generated by monitoring applications, which is still one the predominant use cases of wireless sensor networks.} We simulate 24 hours of network operations and 95\% confidence intervals are computed by replicating each simulation five times with different random seeds. Confidence intervals are shown as error bars in the following diagrams. 

\begin{figure}[tb] 
\centering
    \subfloat[Mobile node (MN) in a corridor\label{fig:corridor}]{%
      \includegraphics[trim={1.5cm 0cm 16cm 11cm},clip,angle=0,width=0.25\textwidth]{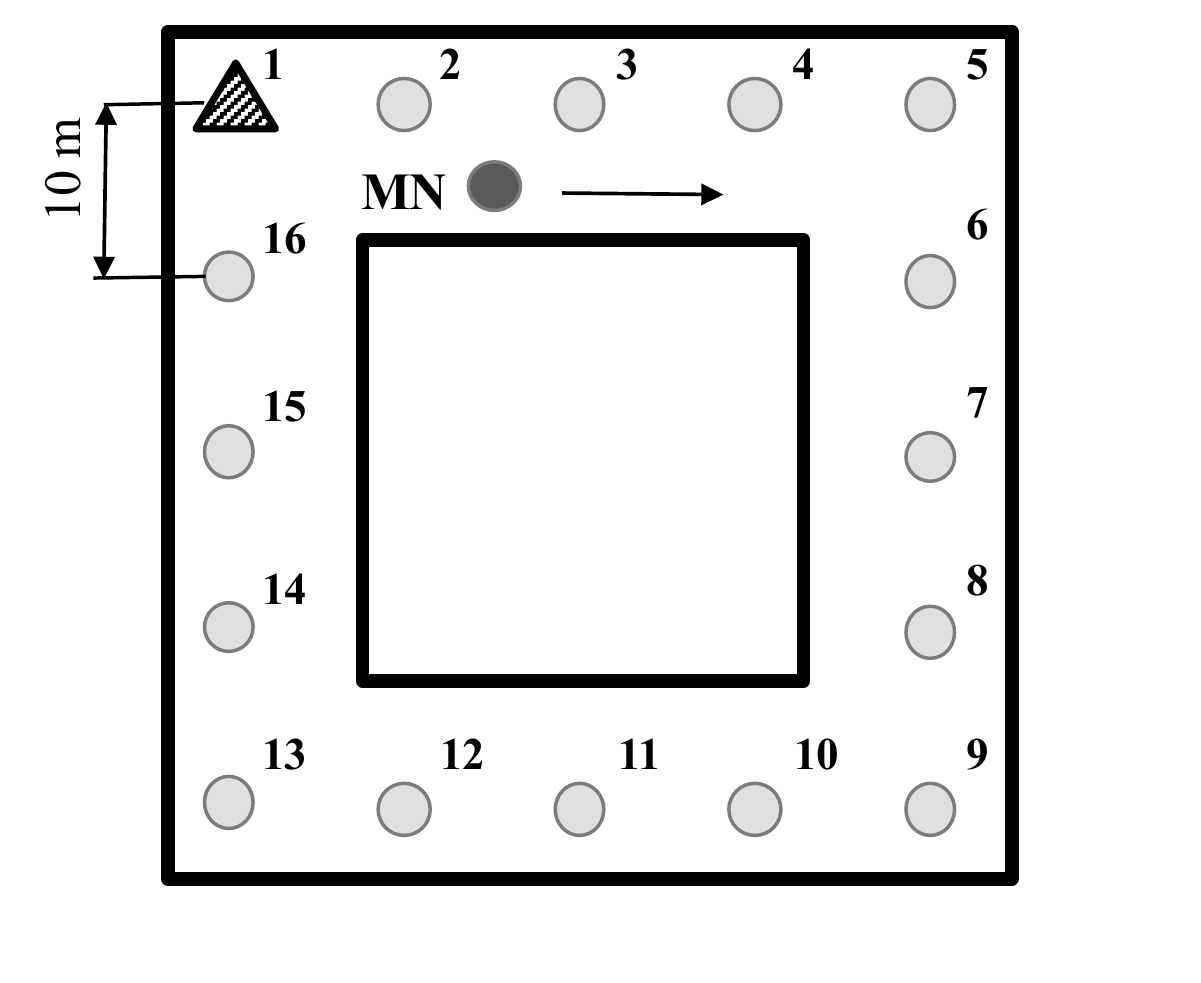}
    }
%
%
    \subfloat[Mobile obstacles\label{fig:obstacle}]{%
      \includegraphics[trim={2cm 0cm 17cm 12cm},clip,angle=0,width=0.25\textwidth]{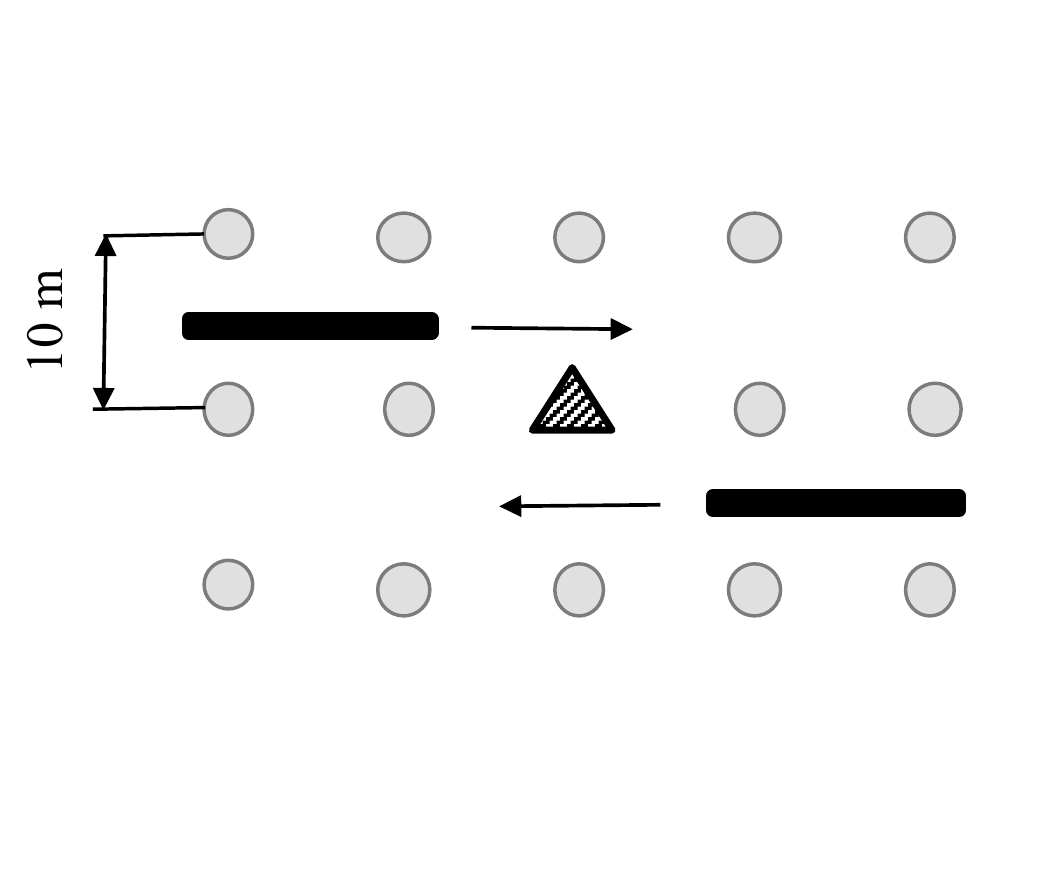}
      }
    \subfloat[Mobile node (MN) in a grid\label{fig:grid}]{%
      \includegraphics[trim={1cm 0cm 12cm 7cm},clip,angle=0,width=0.35\textwidth]{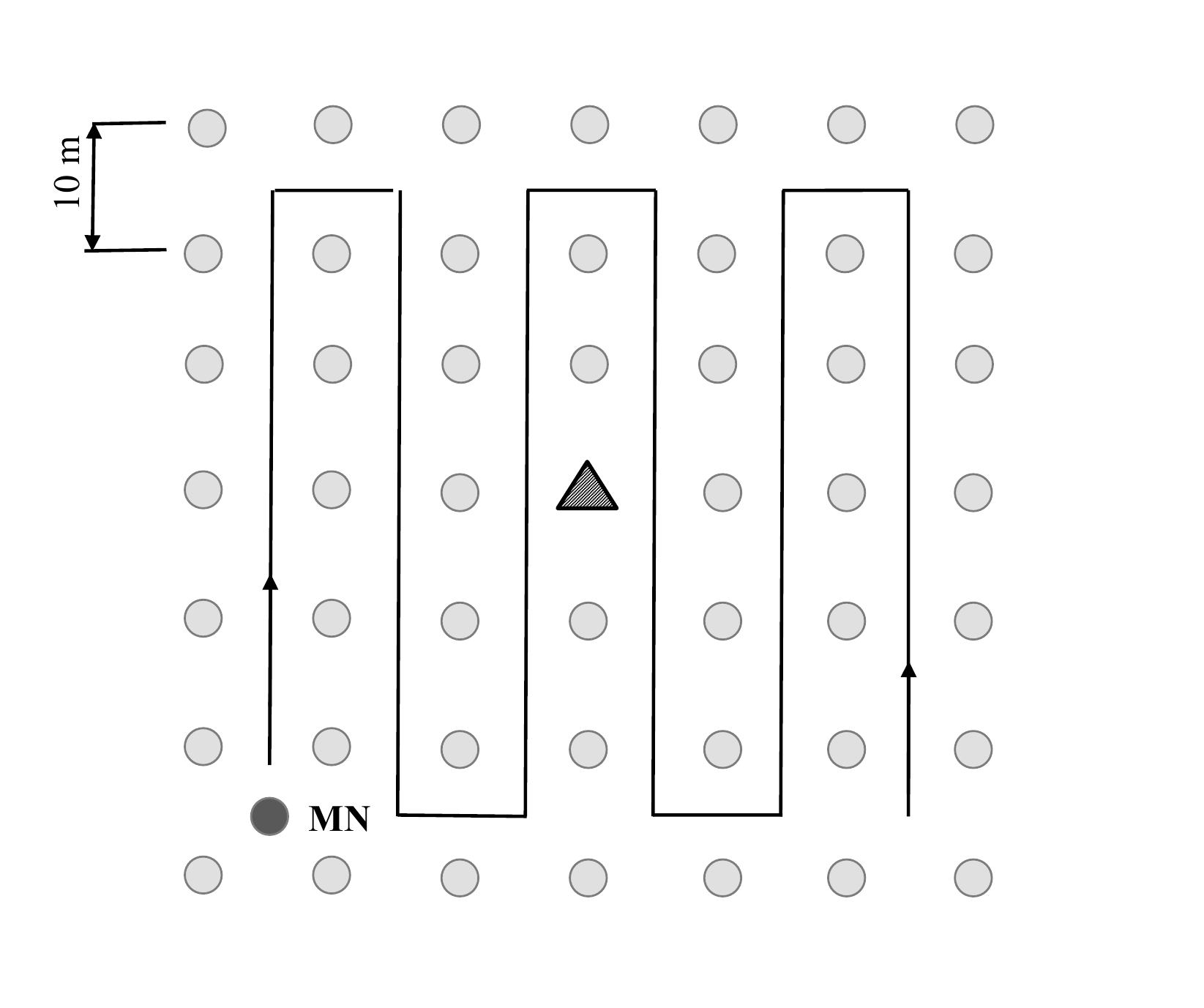}
      }
\caption{Simulation scenarios (the triangle is the root node).}
\label{fig:scenarios}
\vspace{-0.3cm}
\end{figure}

To evaluate the proposed scheme, we consider three network scenarios. The first one is depicted in Figure~\ref{fig:corridor}, and it exemplifies a corridor monitored with fixed and mobile nodes. Specifically, 16~sensor nodes are deployed following a square layout at a distance of 10~meters each. Then, a mobile node moves at a constant speed $v$ from one corner to the following one, and every time it reaches the location of a fixed node it pauses for $p$ minutes. The second scenario is depicted in Figure~\ref{fig:obstacle}, and it exemplifies sensor nodes deployed in a challenged industrial environment (e.g., an assembly line) with large moving obstacles that can impair wireless communications. Specifically, we have three parallel rows of 5 sensor nodes each with one large obstacle that moves from the left to the right corners and back, and another large obstacle that moves in the opposite direction. We assume that each obstacle moves to the next node in the row and remain fixed for a time $p$. Furthermore, each obstacle is able to completely filter out wireless transmissions between adjacent nodes. \added{The last scenario is depicted in Figure~\ref{fig:grid}, and it exemplifies a dense sensor deployment. Specifically, 50~sensors are deployed in a regular grid layout and a mobile node moves at a constant speed following the trajectory shown in the figure. This last scenario is used to demonstrate the performance of the proposed solution in situations in which the mobile node may have many neighbours with high-quality links to choose while changing the preferred parent.}

\added{It is important to point out that in a network in which there are many mobile nodes, or many nodes experiencing unstable links as in the case illustrated in Figure~\ref{fig:obstacle}, it might be difficult to build an optimal network topology using mRPL. Indeed, mobile nodes are forced to act only as leaf nodes and they cannot forward traffic from other nodes. Thus, mRPL could lead to less efficient, or even disconnected, network topologies if there is a large number of nodes configured as mobile/nomadic. In the network scenario depicted in Figure~\ref{fig:obstacle} we have configured the four corner nodes as mobile in mRPL.}.  
%
%
%
\subsubsection{Results with mobile nodes\label{sec:results_mn}}
\begin{figure}[pt] 
\centering
  \subfloat[PLR\label{fig:corridor_plr}]{%
     \includegraphics[trim={1cm 2.5cm 6cm 18.5cm},clip,angle=0,width=0.45\textwidth]{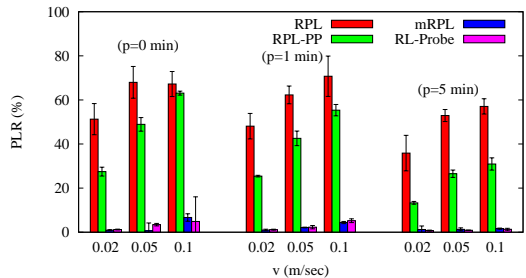}
    }
\\
    \subfloat[Packet overhead\label{fig:corridor_overh}]{%
      \includegraphics[trim={1cm 2.5cm 6cm 18.5cm},clip,angle=0,width=0.45\textwidth]{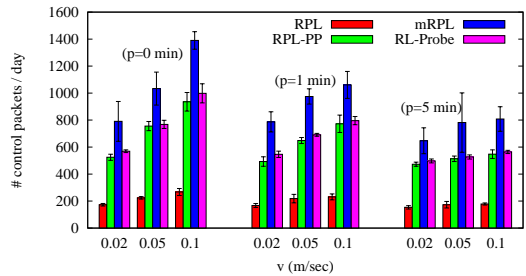}
    }
\\
    \subfloat[\added{Normalised} energy consumption\label{fig:corridor_ec}]{%
      \includegraphics[trim={1cm 2.5cm 6cm 18.5cm},clip,angle=0,width=0.45\textwidth]{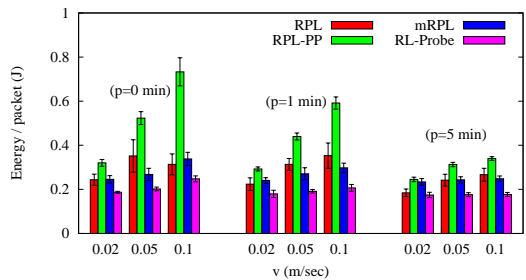}
    }
\caption{Simulation analysis for the \added{scenario shown in Figure~\ref{fig:corridor}} for different speed values and pause times\label{sec:results_corridor}.}
\label{fig:mn_results}
\vspace{-0.3cm}
\end{figure}

\noindent
In this section, we report the results for the scenario illustrated in Figure~\ref{fig:corridor} and Figure~\ref{fig:grid}. 
%
%
%
\paragraph{Corridor topology}
Figure~\ref{fig:corridor_plr} shows the average packet loss rate of the mobile node for various speeds and pause times for the scenario illustrated in Figure~\ref{fig:corridor}. \added{We recall that each time the mobile node reaches the location of a fixed node it stops for $p$ minutes. We refer to this stop period as the pause time of the mobile node.} Important conclusions can be drawn from these results. First, as expected packet loss rates increase when increasing speed and decrease when increasing pause times for all considered schemes. This is due to more frequent handoffs. Secondly, passive monitoring is unable to promptly cope with topology changes and packet loss rates range from 35\% to 65\% in the considered scenarios. Thirdly, unicast-based probing improves RPL ability to detect handoffs but packet loss rates still range from 18\% to 55\%. On the contrary, RL-Probe and mRPL have similar performance and they dramatically improve communication reliability, with packet loss rates that now range from 2\% to 12\%. An in-depth explanation of the root cause of such improvement is provided later in this section. 

Figure~\ref{fig:corridor_overh} shows the protocol overhead in terms of the total number of RPL control messages sent during 24 hours. Clearly, there is a significant increase in packet overhead when active probing is used. As expected, the higher the speed and/or the shortest the pause time (i.e., the faster the network dynamics), the higher the protocol overhead. Interestingly, mRPL generates the highest protocol overhead among the considered routing schemes, while RL-Probe has similar overhead performance as RPL-PP. Analysing more in details the results we found out that the adaptive beaconing of RL-Probe reduces the number of unicast-based probing that are generated with respect to RPL-PP. However, these protocol overhead savings are compensated by the receiver-side probing, which generates trains of consecutive probes. On the other hand, handoff process in mRPL is quite aggressive as it generates long trains of multicast DIS messages to neighbouring nodes. 

It may be argued that an increase in protocol overhead would severely affect the node energy consumption. To verify this conjecture, Figure~\ref{fig:corridor_ec} shows the \added{normalised} energy consumption as estimated by \added{the EnergyTest module provided with }Cooja. \added{We observe that RPL-PP consumes the highest amount of energy per successfully transmitted packet among the considered protocols. In highly dynamic scenarios (e.g. high speeds) RPL-PP consumes up to 100\% more energy than RL-Probe. Interestingly, RPL and mRPL achieve similar normalised energy consumptions, while RL-Probe consumes up to 30\% less energy than the other protocols}. This counterintuitive result can be explained by observing that retransmissions have a great impact on the energy consumption. Thus, avoiding the use of lossy links can balance the additional energy consumption due to active probing.

\begin{figure}[pt] 
\centering
    \subfloat[RPL\label{fig:mn_rpl}]{%
      \includegraphics[trim={2cm 6cm 1cm 6cm},clip,angle=0,width=0.45\textwidth]{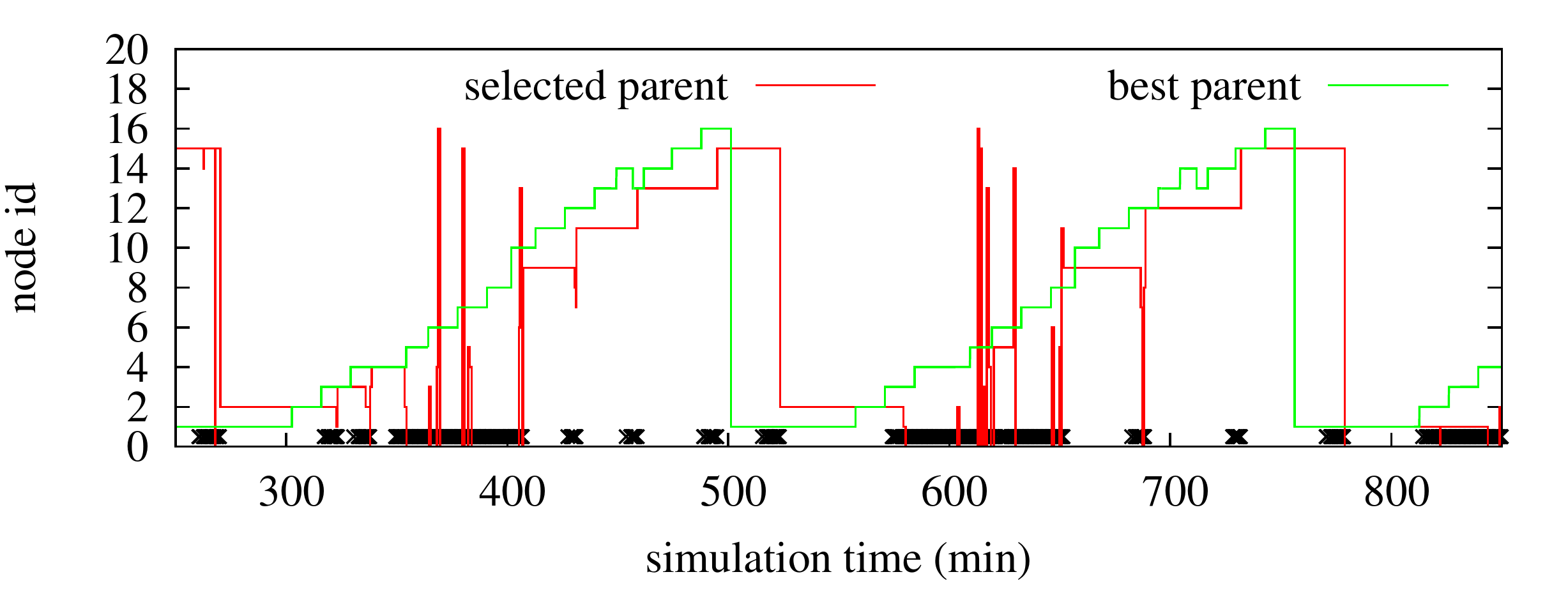}
    }
%
\\
    \subfloat[RPL-PP\label{fig:mn_rpl-pp}]{%
      \includegraphics[trim={2cm 6cm 1cm 6cm},clip,angle=0,width=0.45\textwidth]{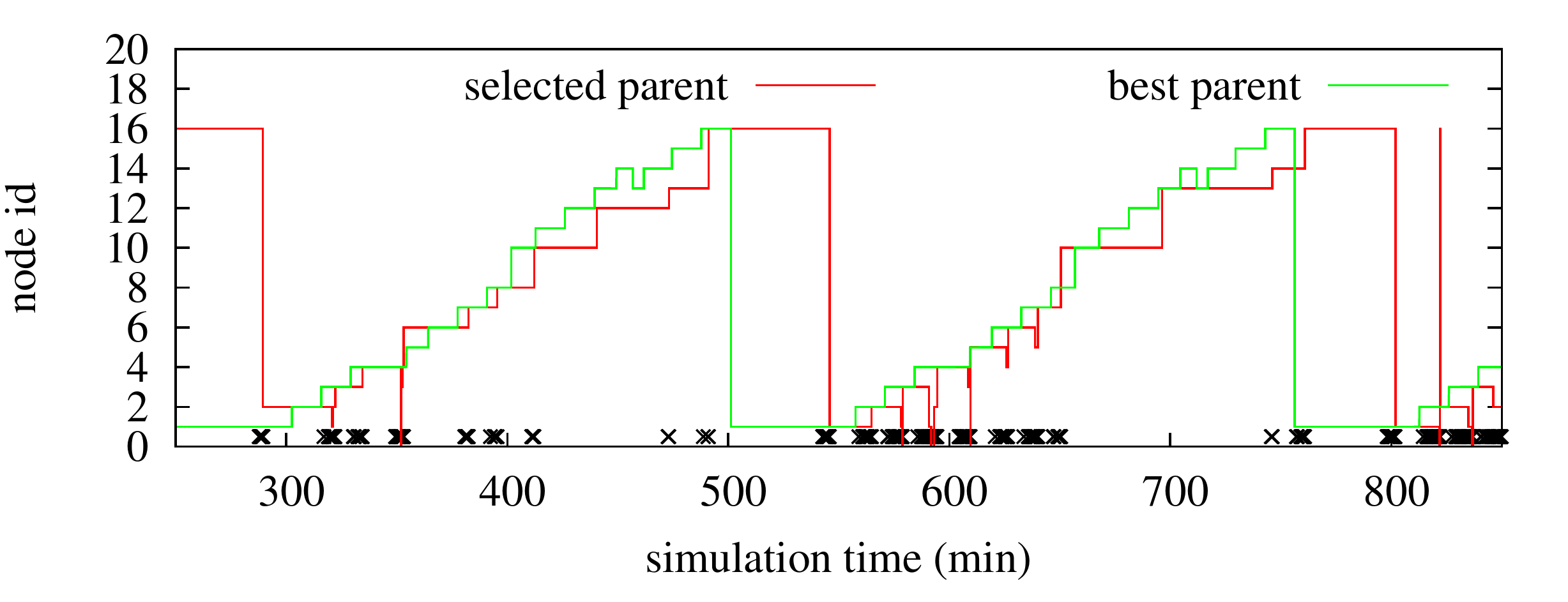}
    }
    %
\\
    \subfloat[mRPL\label{fig:mrpl-pp}]{%
      \includegraphics[trim={2cm 6cm 1cm 6cm},clip,angle=0,width=0.45\textwidth]{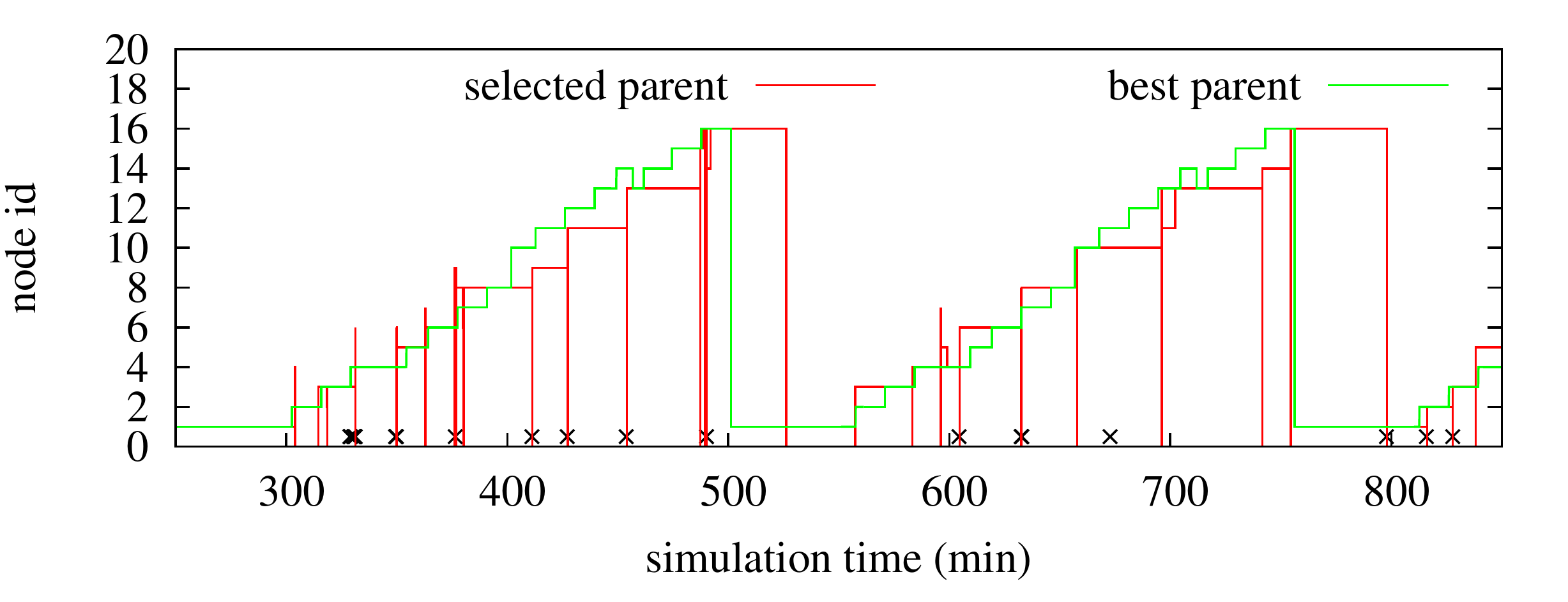}
    }
%
\\
    \subfloat[RL-Probe\label{fig:mn_rl-probe}]{%
      \includegraphics[trim={2cm 6cm 1cm 6cm},clip,angle=0,width=0.45\textwidth]{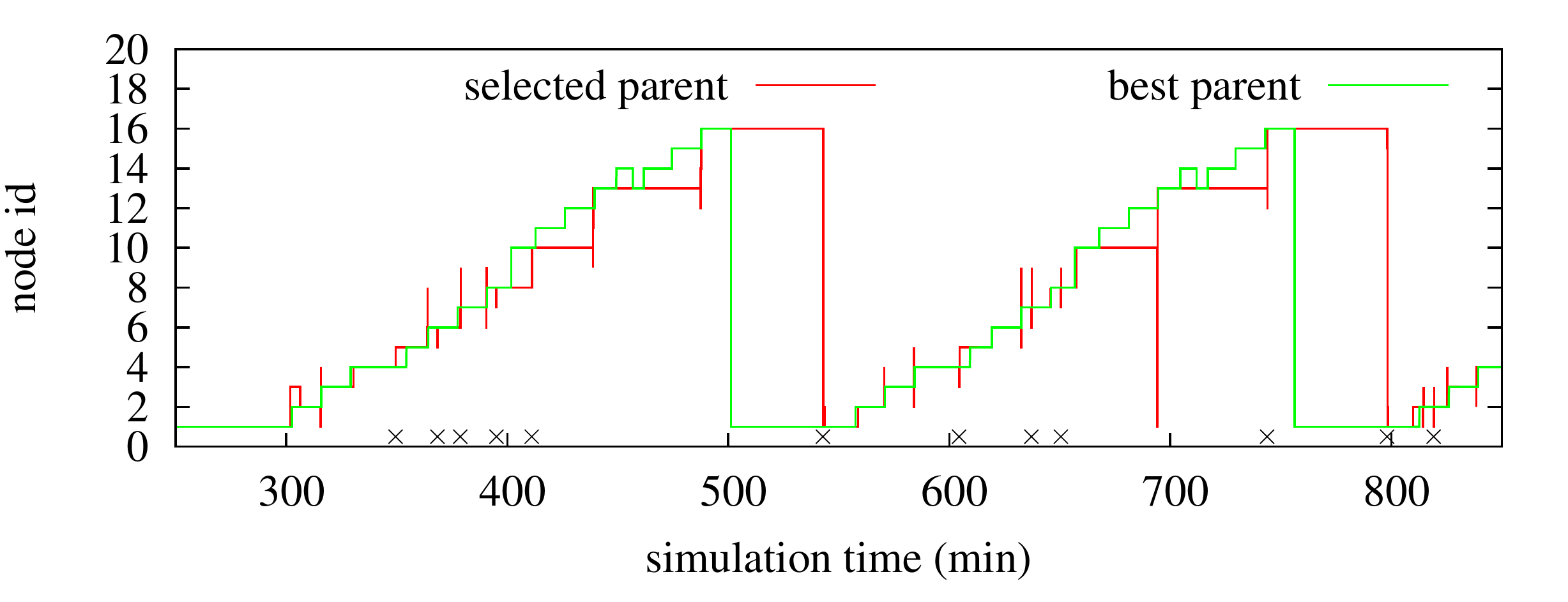}
    }
\caption{Handoff events and packet losses (crosses) for MN when $p=0.01$~m/sec and $p=5$~minutes.}
\label{fig:mn_handoffs}
\vspace{-0.3cm}
\end{figure}
\begin{figure}[pt] 
\centering
    \subfloat[PLR\label{fig:grid_plr}]{%
      \includegraphics[trim={1cm 2.5cm 6cm 18.5cm},clip,angle=0,width=0.45\textwidth]{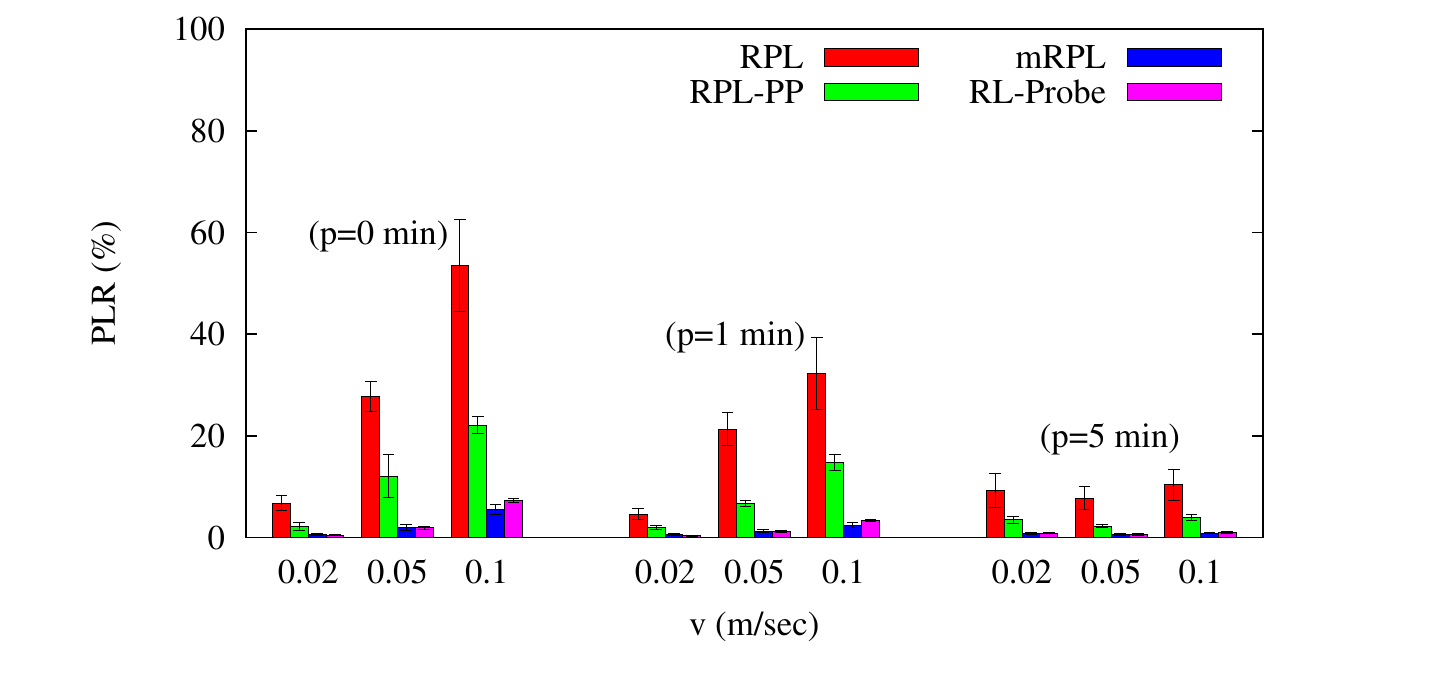}
    }
\\
    \subfloat[Packet overhead\label{fig:grid_overh}]{%
      \includegraphics[trim={1cm 2.5cm 6cm 18.5cm},clip,angle=0,width=0.45\textwidth]{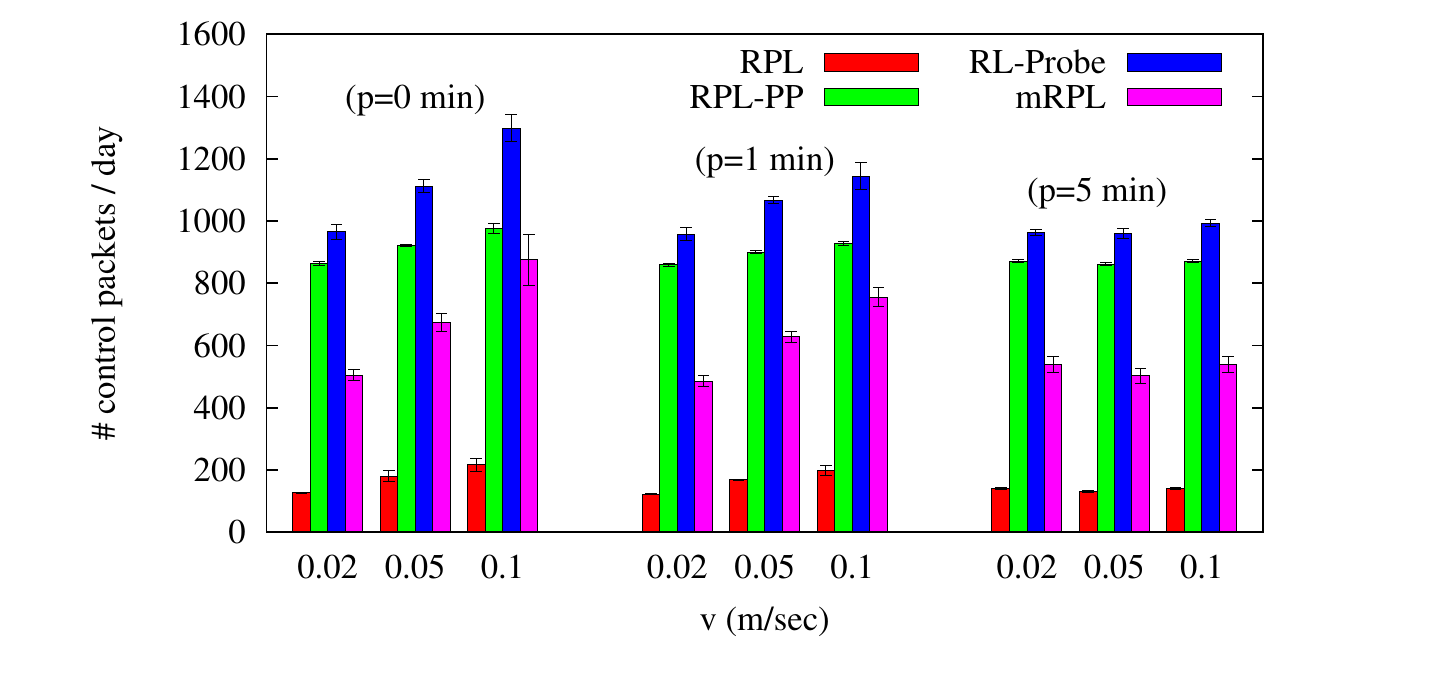}
    }
\\
    \subfloat[Normalised energy consumption\label{fig:grid_ec}]{%
      \includegraphics[trim={1cm 2.5cm 6cm 18.5cm},clip,angle=0,width=0.45\textwidth]{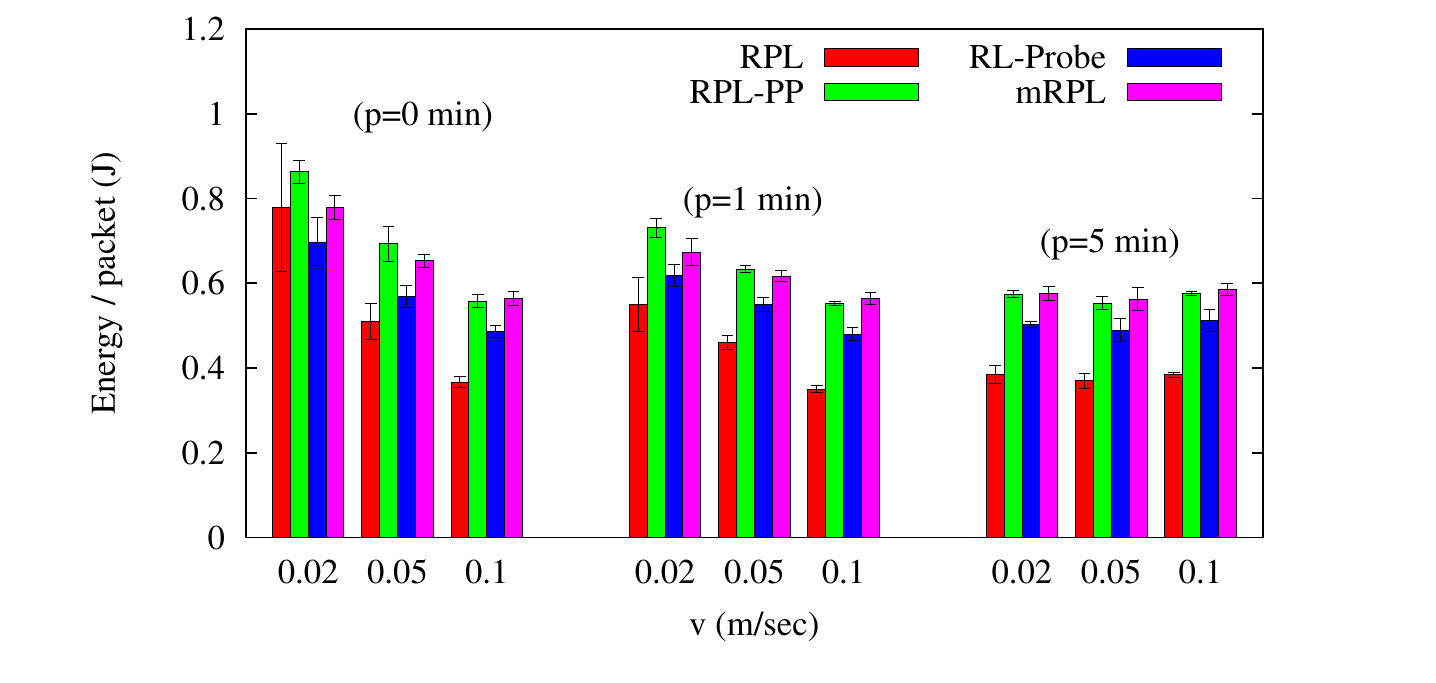}
    }
\caption{Simulation analysis for the \added{scenario shown in Figure~\ref{fig:grid}} for different speed values and pause times\label{sec:grid_results}.}
\label{fig:grid_results}
\vspace{-0.3cm}
\end{figure}

To explain more in details the key advantages of RL-Probe, Figures~\ref{fig:mn_handoffs} show the sequence of handoff events and packet losses for the case $v=0.02$~m/sec and $p=5$~minutes (the best case for both RPL and mRPL). The results indicate that standard RPL with passive link monitoring frequently changes the preferred parent, and it rarely selects the best parent. \added{We define as the best parent the node that would be selected as preferred parent for uplink traffic if each node had a perfect knowledge of the link quality}. RPL-PP is able to follow more closely the best parent. However, handoffs are mainly triggered by packet losses and round-robin periodic probing in RPL-PP causes a burst of packet losses before discovering the new optimal parent. A similar issue is also observed in mRPL, which triggers the discovery phase after a packet loss. Furthermore, the handoff delays in mRPL also cause short disconnections of the mobile node (preferred parent id equal to 0). On the contrary, \emph{the analysis of link trends allows RL-Probe to anticipate changes of link characteristics} and to timely switch to a better preferred parent. Finally, it is interesting to note that a higher number of packet losses occurs when the mobile node is close to the sink. This can be explained by observing that an inaccurate ETX estimation of link quality to neighbours has a greater effect on the rank computation of nodes close to the sink than on nodes far from the sink. 
%
%
%
\paragraph{Grid topology}
\added{Figure~\ref{fig:grid_plr} shows the average packet loss rate of the mobile node for various speeds and pause times for the scenario illustrated in Figure~\ref{fig:grid}. The main noticeable difference with respect to related results shown in Figure~\ref{fig:corridor_plr} is that PLR values are lower for the grid topology than the corridor topology. For instance, with legacy RPL packet loss rates range from 5\% to 55\% in the considered scenarios, and not from 18\% to 65\% as in the corridor topology. This is due to the fact the grid topology has a higher node density and each node has neighbours with good and intermediate quality links.  Nevertheless, general trends are confirmed. First, packet losses increase as the node speed increases or the pause times decreases. Second, conventional RPL experiences packet losses that are one order of magnitude higher than the other schemes. Finally, RL-Probe performs mostly the same as mRPL in all considered cases, and they both outperform RPL-PP.}

\added{Figure~\ref{fig:grid_overh} shows the protocol overhead in terms of the total number of RPL control messages sent during 24 hours. Conventional RPL has a packet overhead that is from three to four times lower than the other schemes that uses active probing. As expected, the higher the speed and/or the shortest the pause time (i.e., the faster the network dynamics), the higher the protocol overhead. As also shown in Figure~\ref{fig:corridor_overh} mRPL generates the highest protocol overhead among the considered routing schemes, while RL-Probe has similar overhead performance as RPL-PP.}

\added{Figure~\ref{fig:grid_ec} shows the normalised energy consumption as estimated by the EnergyTest module provided with Cooja. We observe that RPL consumes the least amount of energy per successfully transmitted packet as it uses only passive techniques for link quality monitoring. RPL-PP and mRPL behaves similarly in all considered scenarios while mRPL consumes up to 20\% less energy than RPL-PP and mRPL. }
%
%
%
\subsubsection{Results with mobile obstacles}
\begin{figure}[pt] 
\centering
    \subfloat[PLR (\added{bars represent the average values and filled squares the maximum value})\label{fig:obstacle_plr}]{%
      \includegraphics[trim={1cm 3cm 7cm 18cm},clip,angle=0,width=0.45\textwidth]{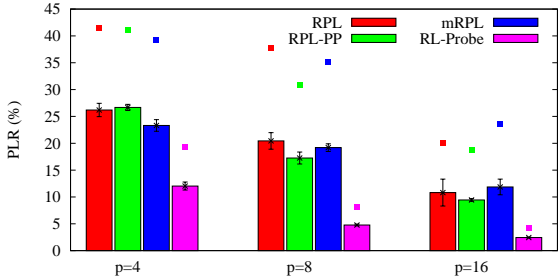}
    }
%
\\
    \subfloat[Packet overhead\label{fig:obstacle_overh}]{%
      \includegraphics[trim={1cm 3cm 7cm 18cm},clip,angle=0,width=0.45\textwidth]{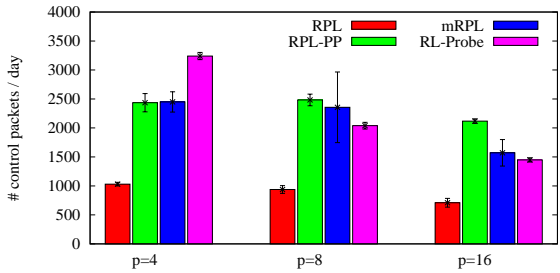}
    }
    %
\\
    \subfloat[\added{Normalised} energy consumption\label{fig:obstacle_ec}]{%
      \includegraphics[trim={1cm 3cm 7cm 18cm},clip,angle=0,width=0.45\textwidth]{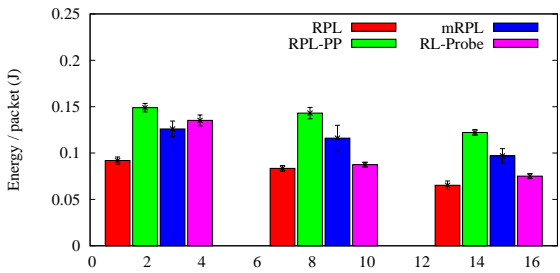}
    }
\caption{Simulation analysis for the \added{scenario shown in Figure~\ref{fig:obstacle}} for different pause times.\label{fig:obstacle_results}}
\vspace{-0.3cm}
\end{figure}

\noindent
In this section, we report the results for the scenario illustrated in Figure~\ref{fig:obstacle}, where network topology changes are due to variations of link conditions caused by mobile obstacles and not handoffs. Figure~\ref{fig:obstacle_plr} shows the average (bars) ad maximum (squares) PLR of \emph{all} nodes for different pause times ($p=4,8,16$~minutes). Results indicate that RL-Probe achieves a three-fold decrease of both average and peak PLRs with respect to the other considered schemes, including mRPL. On the contrary, mRPL performs similarly to RPL and RPL-PP. Several factors contribute to mRPL inefficient behaviour. First, the hysteresis margin in mRPL assumes that  the transitional region of links is quite wide~\cite{Fotouhi2014_tmc}. However, this decreases the ability of mRPL to detect sudden changes of link quality that occur within the transitional region. Furthermore, if the quality of the link to the preferred parent is stable mRPL does not trigger discovery phases, which are needed to quickly detect if the quality of the links to neighbouring nodes is suddenly improved (e.g., because an obstacle has moved). Figure~\ref{fig:obstacle_overh} shows the protocol overhead in terms of RPL control messages. We can observe that RL-Probe rapidly limits protocol overhead as the network conditions become less variable (i.e., pause times increase). RL-Probe generates higher protocol overhead than mRPL only for $p=4$~minutes, but lower overhead for $p=8$ and $p=16$~minutes. The same trend can be observed also for the total energy consumption (see Figure~\ref{fig:obstacle_ec}). Summarising, in case of link quality variability due to changes in network conditions RL-Probe outperforms mRPL in terms of communication reliability with similar network overhead and energy waste.

%
%
%
%
%
%
%
\subsection{Experimental Evaluation}
\begin{figure}[t]
\centering
\includegraphics[width=0.4\textwidth]{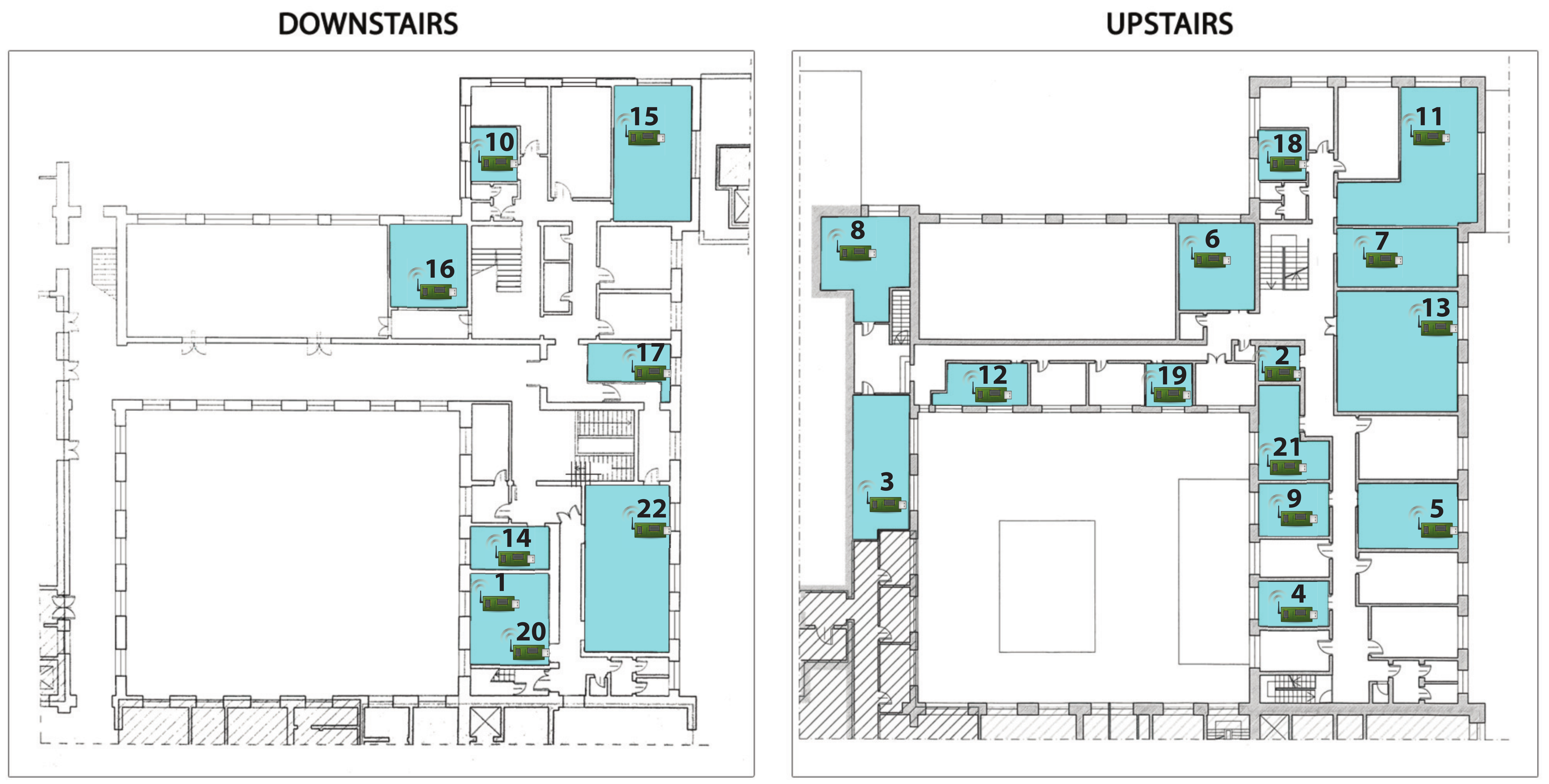}
\caption{Map of the testbed\label{fig:testbed}}
\vspace{-0.3cm}
\end{figure}

\noindent
In this section, we report the results obtained from real experiments conducted in an indoor IoT testbed \cite{pewasun16_vallati}. Specifically, our low-power wireless network is composed of 23 wireless sensor nodes deployed in office spaces, student labs, and corridors on two floors in the Department of Information Engineering of the University of Pisa. Figure~\ref{fig:testbed} shows the layout of the testbed. Sensor nodes are TelosB motes, equipped with an MSP430 micro-controller that can run a wide range of Operating Systems for sensors. Thus, the same Contiki code used for Cooja simulations is loaded on the testbed. IEEE802.15.4 connectivity is provided through the cc2420 wireless chip equipped with an external 5dBi antenna. The maximum number of active links in the network is 178. Table~\ref{tab:links} reports the main percentiles of the average ETX across all links, as measured by unicast-based probing without any data or control traffic in the network.

\begin{figure}[pt] 
\centering
  \subfloat[PLR (bars represent the average values and filled squares the maximum value)\label{fig:exp_static_plr}]{%
     \includegraphics[trim={1cm 1cm 7cm 18cm},clip,angle=0,width=0.45\textwidth]{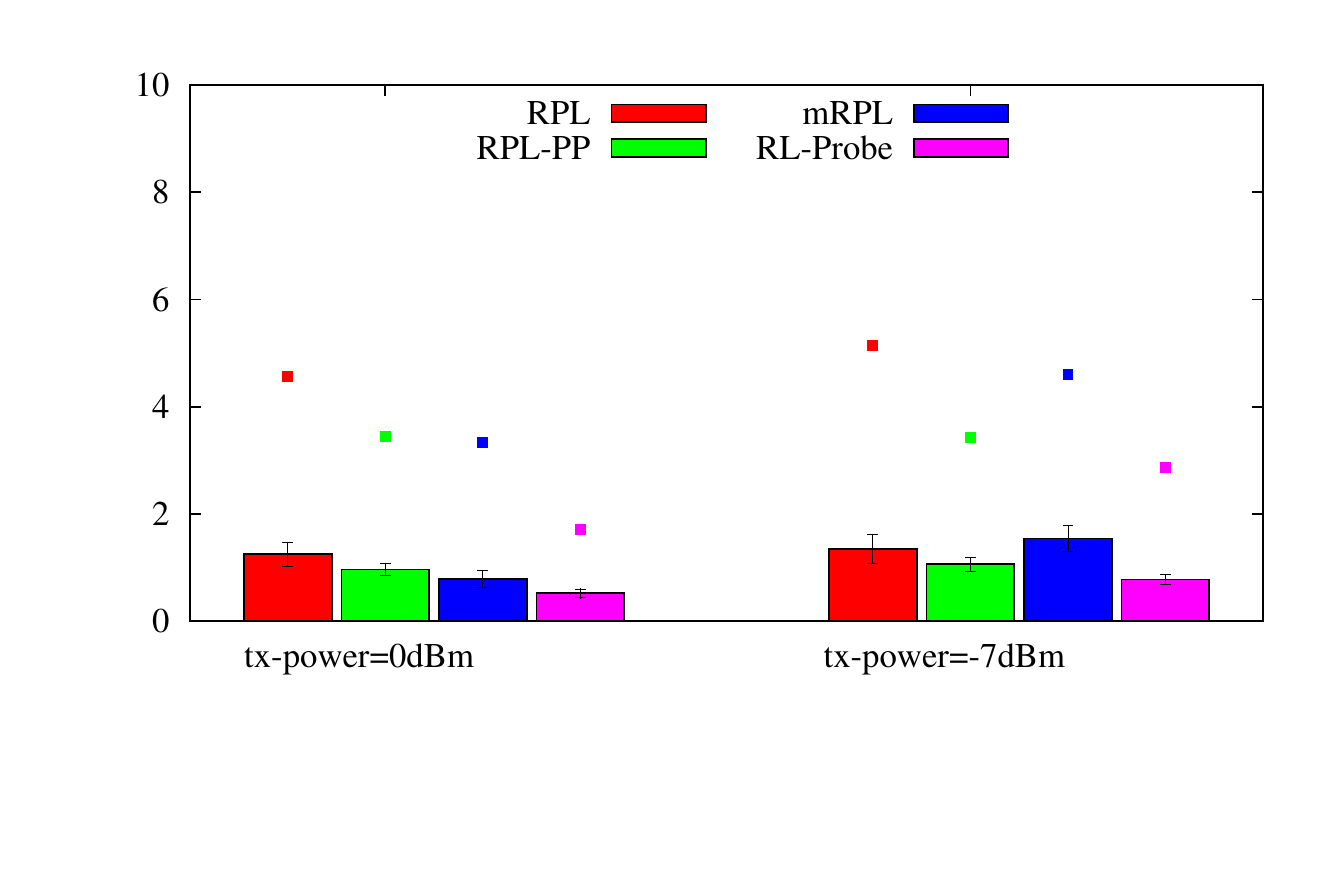}
    }
\\
    \subfloat[Normalised energy consumption\label{fig:exp_static_ec}]{%
     \includegraphics[trim={1cm 1cm 7cm 18cm},clip,angle=0,width=0.45\textwidth]{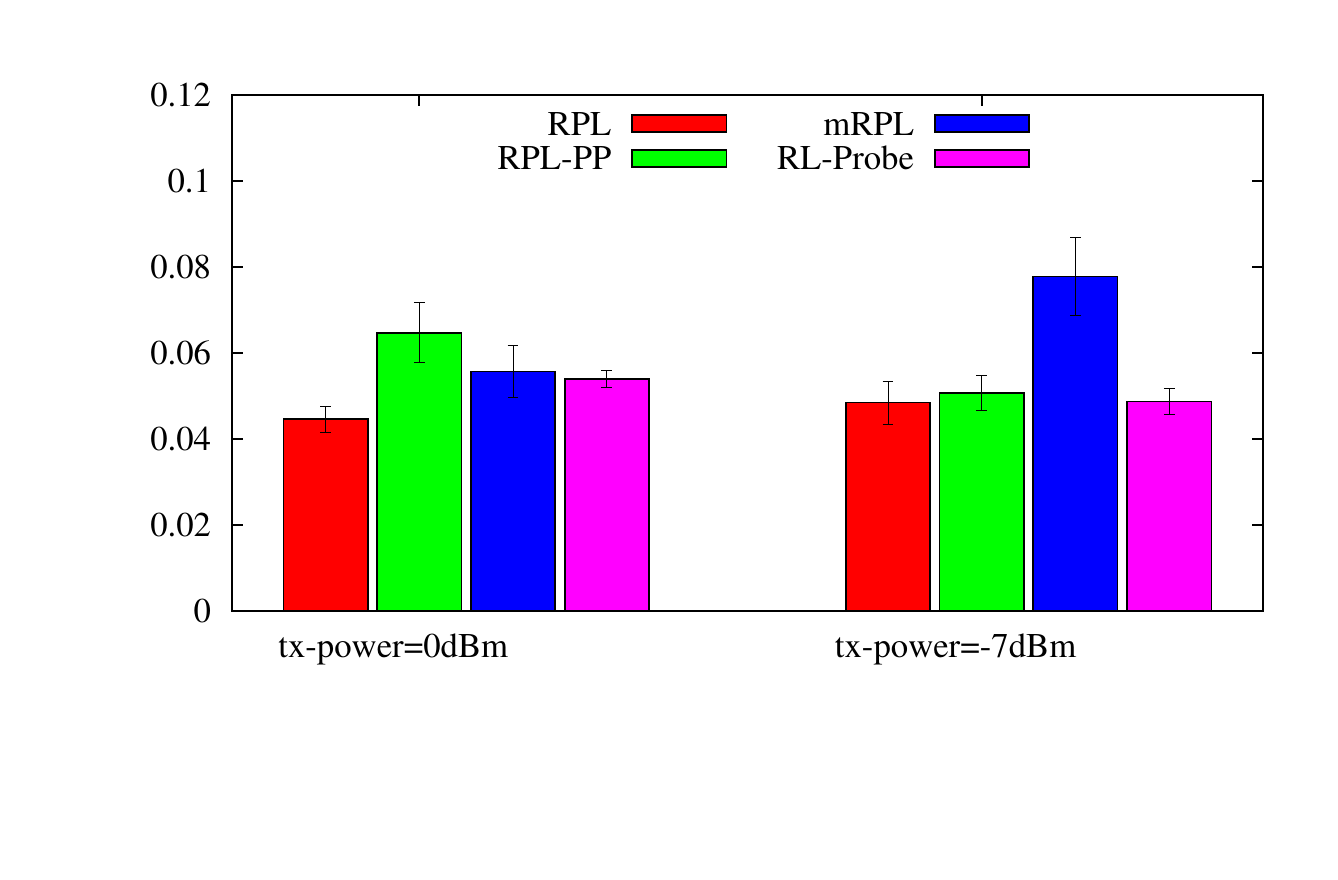}
    }
\caption{\added{Experimental analysis of the static scenario for different transmission powers.}\label{fig:exp_static}}
\vspace{-0.3cm}
\end{figure}

 \added{The first set of results is obtained considering a static scenario in which there are no mobile nodes. However, we emphasise that our testbed is deployed in a dynamic environment and the experiments have not been run at special times to avoid interference. Thus, our testbed is susceptible to changes in radio channel conditions due to interference (e.g., from other 802.15.4 radios and from 802.11 radios), and this interference is highly time-varying. Furthermore, we replicated each test using two radio transmission powers: 0dBm and -7dBm. The first value is the maximum transmission power that is supported by the CC2420 RF transceiver, which clearly maximises the network density and the number of high-quality links. The latter value is used to evaluate the performance in a configuration in which links with intermediate quality also exist. Note that farther reducing the transmission power may lead to a partitioned network topology. Finally, all nodes are configured to generate a CBR traffic consisting of 40-bytes UDP messages sent every minute to node 1. The underlying MAC protocol is CSMA and ContikiMAC is used as RDC layer.  Figure~\ref{fig:exp_static_plr} shows the average packet loss ratio and energy consumption measured for the different RPL variants during experiments that last three hours. The results confirm that adding active probing techniques is beneficial to improve the routing reliability because it enables faster routing adaptation to channel fluctuations. In addition, in the analysed scenarios RPL does not experience high PLR values. Reducing the transmission power has only a small impact on the overall PLRs. Figure~\ref{fig:exp_static_ec} shows the normalised energy consumption for the different strategies. The results demonstrate that energy consumptions of RPL variants are basically equivalent}. 
\begin{table}
\caption{$p$-th percentiles of the average ETX of the wireless links in the testbed.\label{tab:links}}
\vspace{0.2cm}
\centering 
\begin{tabular}{| l | c |c | c | c | c |}
\hline
 & \multicolumn{5}{c|}{$p$ value} \\
\cline{2-6}
 & 10\%&  25\% & 50\% & 75\%& 90\% \\
\hline
 Average ETX & 1.0 & 1.0 & 1.025 & 1.245 & 2.776 \\
\hline
\end{tabular}
\end{table}

\begin{figure}[tb] 
\centering
\includegraphics[trim={0cm 0cm 0cm 0cm},clip,angle=0,width=0.35\textwidth]{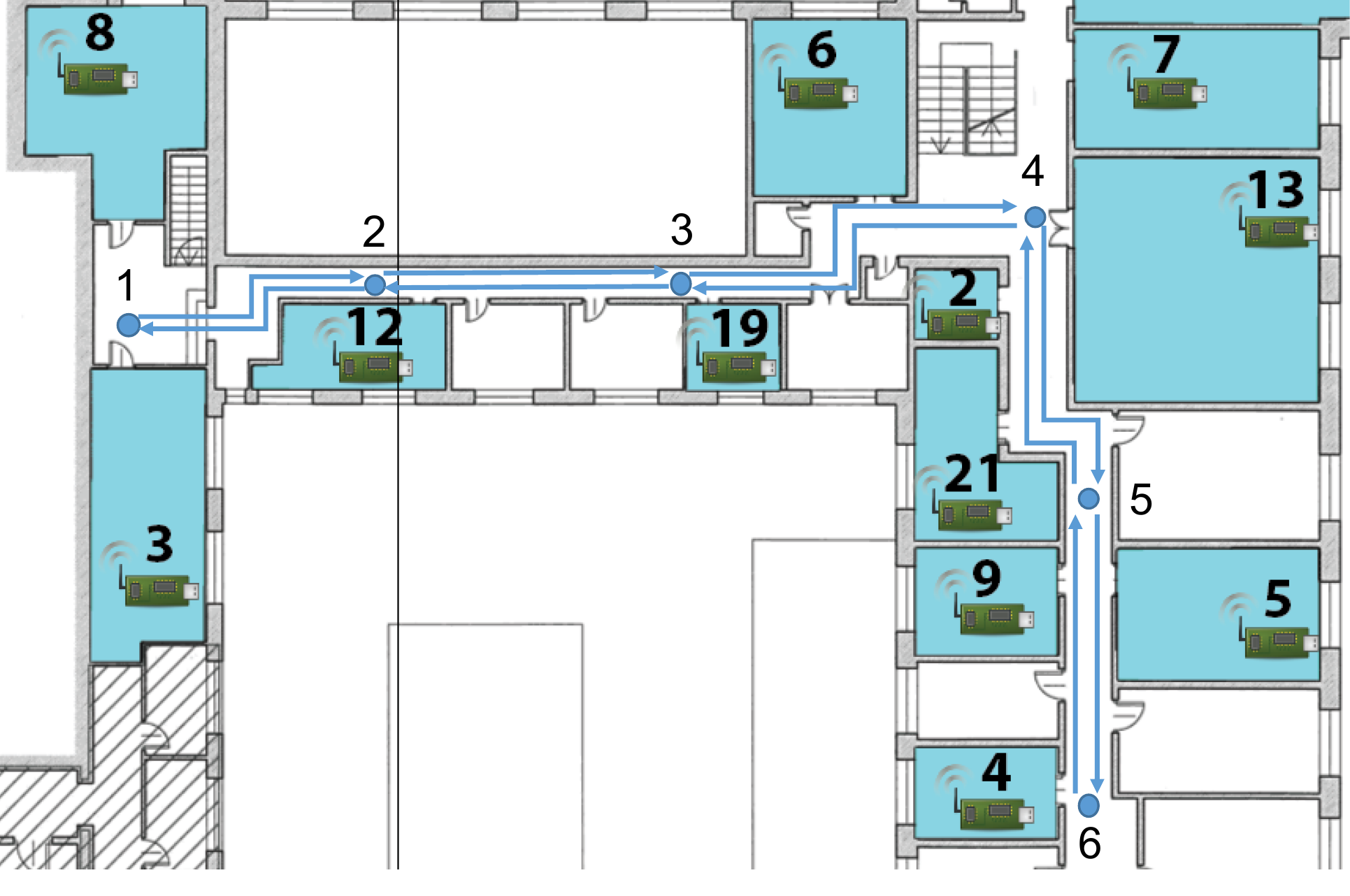}
\caption{Trajectory followed by the mobile node during the mobility experiments.\label{fig:mobile_path}}
\vspace{-0.3cm}
\end{figure}

The second set of results is obtained in a scenario in which there is a mobile sensor node. Figure~\ref{fig:mobile_path} illustrates the trajectory of this mobile node. Specifically, after an initial set-up phase of 5 minutes, the node moves at 0.5 m/s from one specified point to another, pausing at each location for 2 minutes.  The path is covered round-trip, i.e. from point 1 to 6 and then back from 6 to 1. Traffic is only originated by the mobile node towards node number 1, which is selected as sink and RPL root node. A CBR traffic is employed i.e. a 40-byte UDP message is emitted every 10~seconds. \added{In these tests, the radio transmission power is set to -7 dBm to use a sparser network topology}. This guarantees that each movement of the mobile node results into a change of the parent node. Each experiment lasts 30 minutes.

\begin{figure}[tb] 
\centering
\includegraphics[trim={1cm 1cm 7cm 18cm},clip,angle=0,width=0.45\textwidth]{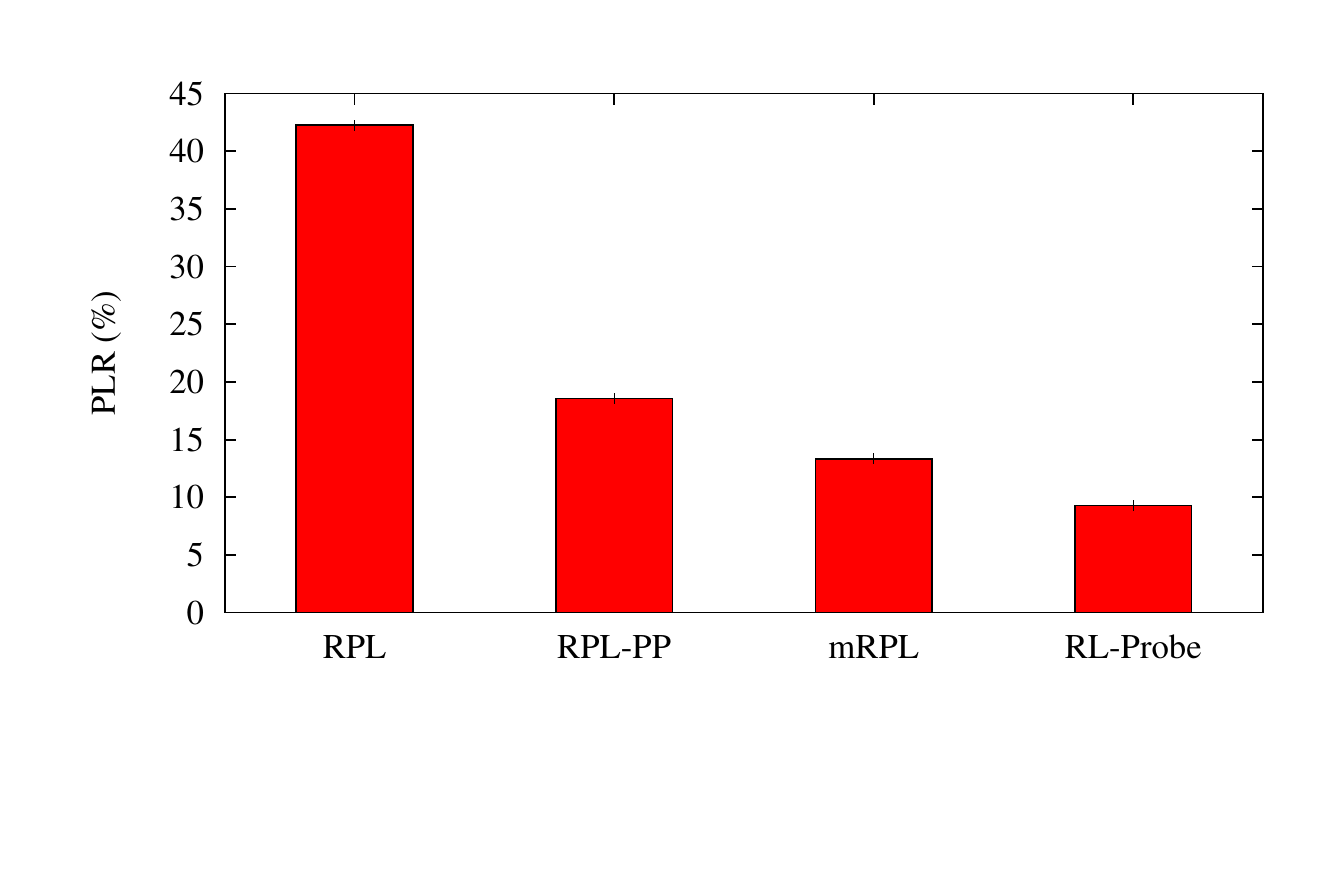}
\caption{Experimental analysis: average packet loss for the mobile scenario.\label{fig:exp_mobile_pkt_loss}}
\vspace{-0.3cm}
\end{figure}

Figure~\ref{fig:exp_mobile_pkt_loss} shows the average packet loss experienced by the mobile node with different strategies.  We can observe that RL-Probe slightly outperforms mRPL, which confirms the effectiveness of multicast probing in obtaining a rapid assessment of the link quality when multiple neighbours appear/disappear at the same time. On the other hand, standard RPL experiences many packet losses every time the mobile node changes its location, due to the lack of an active strategy for LQE. RPL-PP achieves better performance than basic RPL but it is less efficient than both mRPL and RL-Probe. This can be explained by considering that unicast probing assesses links individually and therefore more time is needed to discover a better preferred parent when moving.  

\begin{figure}[tb] 
\centering
\includegraphics[trim={1cm 1cm 7cm 18cm},clip,angle=0,width=0.45\textwidth]{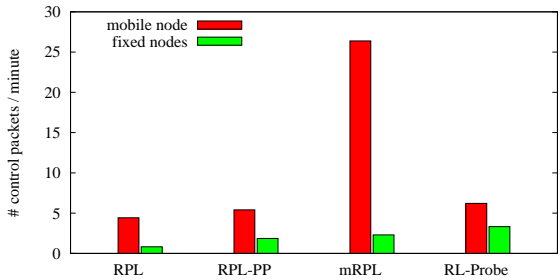}
\caption{Experimental analysis: control message overhead for the mobile scenario.\label{fig:exp_mobile_overhead}}
\vspace{-0.3cm}
\end{figure}

Figure~\ref{fig:exp_mobile_overhead}, instead, quantifies the overhead produced by each strategy. Specifically, the figure reports the average number of RPL control packets (including both probe and response packets) per second generated by the nodes in the network. We distinguish between the overhead generated by static nodes and the overhead generated by the mobile node. As expected, legacy RPL is the strategy characterised by the least overhead, as nodes only transmit RPL control packets for topology discovery without any probe packet. RPL-PP, instead, shows a slight increase in the overhead as a light unicast probe traffic is employed. Both mRPL and RL-Probe are characterised by the highest overhead due to the active probe traffic generated by each node. However, the overhead generated by the mobile node using mRPL is four times the overhead provided by the same mobile node when using RL-Probe. This clearly shows that the responsiveness of mRPL to node mobility is obtained at the cost of introducing frequent probing. On the contrary, RL-Probe does not penalise the mobile node, who requires a minimal additional overhead with respect to fixed nodes to detect link failures. 

\begin{figure}[tb] 
\centering
\includegraphics[trim={1cm 1cm 7cm 18cm},clip,angle=0,width=0.45\textwidth]{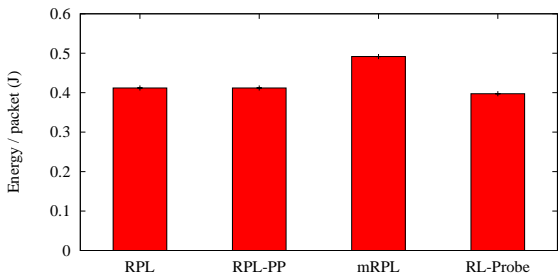}
\caption{Experimental analysis: normalised energy consumption for the mobile scenario.\label{fig:exp_mobile_ec}}
\vspace{-0.3cm}
\end{figure}

\added{Finally, Figure~\ref{fig:exp_mobile_ec} shows the normalised energy consumption for each strategy. Interestingly, we can observe that the energy consumed per successfully transmitted packet by each strategy is equivalent. This can be explained by considering that PLR for standard RPL is higher than the one of the other schemes and this compensates for the overhead increase. Since mRPL generates the highest overhead, it is also characterised by the highest normalised energy consumption.}
%

%
%
%
\section{Conclusion}\label{sec:conclusion}
\noindent
In this paper, we have proposed RL-Probe, link quality estimation strategy for RPL-based WSNs. RL-Probe employs synchronous and asynchronous monitoring schemes to maintain up-to-date information on link quality towards the neighbours and react to sudden topology changes. RL-Probe achieves a trade-off between a low probing overhead and responsiveness to changing network conditions by leveraging on a lightweight reinforcement learning technique to control the active probing operations. This is crucial to minimise energy consumptions of tiny, resource-constrained devices. Furthermore, we have integrated our solution in the RPL implementation that is included in the Contiki operating system for embedded devices. A performance evaluation based on both simulations and real-world experiments has been carried out, demonstrating how the proposed approach guarantees better performance with respect to state-of-the-art LQE techniques for RPL. In particular, results show that the proposed approach does not only properly react to link quality variations, but it is also effective to handle topology variations due to mobility. 

\added{As future work, we plan to investigate how to improve RL-Probe performance in interference-limited scenarios in which link variations are due to external interference sources. One possible approach is to leverage opportunistic communications and to design a LQE techniques for cognitive radio~\cite{2016_RAWAT}. Furthermore, RL-Probe can be extended to cater for more efficient learning policies than the greedy approach. Finally, RL-Probe is designed under the assumption that links are symmetric, thus exploiting eventual link asymmetry  links is an open issue. A possible solution to identify the link asymmetry would be to measure the quality of each link in both directions of the link, as in~\cite{Kim2009_ton}, or to leverage cooperative approaches as in~\cite{Sang2010_tosn}.}
%
%
%
%
\section*{Acknowledgments}\label{sec:ack}
\noindent
We are thankful to the editor and reviewers for their very constructive comments that helped us to significantly improve the paper quality and clarity.  

This work was partially funded by the SIGS project. This project has received funding from the Regione Toscana through the PAR FAS 2007-2013 funds. 
\bibliographystyle{elsarticle-num}
\bibliography{IEEEabrv,biblio,comcom}
\end{document}